\documentclass{tlp}
\usepackage{amssymb}
\usepackage{latexsym}
\usepackage{epsfig}
\usepackage{graphicx}
\usepackage{epic}
\usepackage{rotating}

\newcommand{\HIDE}[1]{}
\newcommand{\FULLS}[1]{}

\newcommand{\FULLCLP}[1]{}

\newcommand{\comment}[1]{}
\newcommand{\extended}[1]{}

\newcommand{\ma}{\mbox{$\:\rhd\:$}}
\newcommand{\maeq}{\mbox{$\:\unrhd\:$}}

\newenvironment{programma}{
\tt \begin{tabbing}program\= {\tt pro}\= clause 
\kill}{\end{tabbing}}

\newcommand{\claus}[2]{\mbox{$#1 \la \tn{#2}{n}$}}

\newcommand{\clausab}{\claus{A}{B}}

\newcommand{\bp}[1]{\mbox{$B_P$}}

\newcommand{\groun}[2]{\mbox{{\it ground}$_{#1}(#2)$}}

\newcommand{\tn}[2]{\mbox{$#1_{1}\:,\ldots\:,#1_{#2}\:$}}

\newcommand{\pre}{\mbox{$Pre$}}
\newcommand{\post}{\mbox{$Post$}}

\newcounter{ProblemCnt}

\newcounter{prgcnt}
\newenvironment{program}{
\tt \begin{tabbing}program\= {\tt pro}\= clause 
\kill}{\end{tabbing}}

\newcommand{\vect}[1]{\mbox{\bf #1}}

\newcommand{\inv}[1]{}

\newcommand{\la}{\mbox{$\:\gets\:$}}
\newcommand{\ra}{\mbox{$\:\to\:$}}

\newcommand{\Ra}{\mbox{$\:\Rightarrow\:$}}

\newcommand{\A}{\mbox{$\ \land\ $}}

\newcommand{\U}{\mbox{$\:\cup\:$}}

\newcommand{\sse}{\mbox{$\:\subseteq\:$}}

\newcommand{\Mo}{\mbox{$\:\models\ $}}

\newcommand{\LL}{\mbox{$\ldots$}}

\newcommand{\dom}{{\it Dom}}

\newcommand{\szkew}[1]{\relax \setbox0=\hbox{\kern -24pt
$\displaystyle#1$\kern 0pt }%
\box0}
{\catcode`\@=11 \global\let\ifjusthvtest@=\iffalse}

\newcommand\newplaintheorem[3]{%
\newtheorem{#1plain}[#3]{#2}
\newenvironment{#1}[1][]{\removebrackets\begin{#1plain}[##1]\rm}{\end{#1plain}}}

\newplaintheorem{example}{Example}{theo}
\newplaintheorem{definition}{Definition}{theo}
\newplaintheorem{theorem}{Theorem}{theo}
\newplaintheorem{lemma}{Lemma}{theo}
\newplaintheorem{proposition}{Proposition}{theo}

\newcommand\inst{\mathit{inst}}
\newcommand\cequals{c=}

\def\comment#1{\marginpar{\scriptsize #1}}

\newcommand{\nat}{{\rm I \hspace{-0.2em} N}}
\newcommand{\codom}{\mathit{Ran}}

\newcommand{\inp}{{\it I}}
\newcommand{\out}{{\it O}}
\newcommand{\vars}{\mathit{Vars}}

\newenvironment{janprogram}{
\tt \begin{tabbing}pro\= {\tt pro}\= clause 
\kill}{\end{tabbing}}

\newcommand\onlyhere[1]{\subsection{#1}}

\newcommand\limitation{}
\newcommand\termination{\subsection{Operational Definition}}
\newcommand\contrived{}
\newcommand\decrease{\subsection{Declarative Characterisation}}
\newcommand\comparison{}
\newcommand\maintheorem{}
\newcommand\examples{\subsection{Examples}}
\newcommand\incompleteness{\subsection{On Completeness of the Characterisation}}

\begin{document}
\bibliographystyle{tlp}

\title{Classes of Terminating Logic Programs} 
\author[D. Pedreschi, S. Ruggieri and J.--G. Smaus]
{DINO PEDRESCHI and  SALVATORE RUGGIERI\\ 
Dipartimento di Informatica, Universit{\`a} di Pisa\\ 
Corso Italia 40, 56125 Pisa, Italy,\\ 
\email{\{pedre,ruggieri\}@di.unipi.it} 
\and
JAN--GEORG SMAUS\thanks{
Supported by the ERCIM fellowship programme.}\\
Institut f{\"u}r Informatik, Universit{\"a}t Freiburg, \\
Georges-K{\"o}hler-Allee 52, 79110 Freiburg im Breisgau, Germany\\
\email{smaus@informatik.uni-freiburg.de}
}

\maketitle

\noindent
{\bf Note:} This article has been published in
\emph{Theory and Practice of Logic Programming}, 
2(3), 369--418, 
\copyright Cambridge
University Press.\\
On July 22, 2002, the following mistake was corrected:
In figure \ref{permute-prog}, the first clause for $\mathtt{insert}$
was \verb!insert([],X,[X])!.\\
\enlargethispage{6ex}

\begin{abstract}
  Termination of logic programs depends critically on the selection
  rule, i.e.~the rule that determines which atom is selected in each
  resolution step. In this article, we classify programs (and queries)
  according to the selection rules for which they terminate. This is a
  survey and unified view on different approaches in the literature.
  For each class, we present a sufficient, for most classes even
  necessary, criterion for determining that a program is in that
  class. We study six classes: a program {\em strongly} terminates if
  it terminates for {\em all} selection rules; a program {\em input
    terminates} if it terminates for selection rules which only select
  atoms that are sufficiently instantiated in their input positions,
  so that these arguments do not get instantiated any further by the
  unification; a program {\em local delay} terminates if it terminates
  for local selection rules which only select atoms that are bounded
  w.r.t.~an appropriate level mapping; a program {\em left-terminates}
  if it terminates for the usual left-to-right selection rule; a
  program {\em $\exists$-terminates} if there exists a selection rule
  for which it terminates; finally, a program has {\em bounded
    nondeterminism} if it only has finitely many refutations. We
  propose a semantics-preserving transformation from programs with
  bounded nondeterminism into strongly terminating programs.
  Moreover, by unifying different formalisms and making appropriate
  assumptions, we are able to establish a formal hierarchy between the
  different classes.
\end{abstract}

\section{Introduction}\label{intro-sec}

The paradigm of logic programming originates from the discovery that a
fragment of first order logic can be given an elegant computational
interpretation. Kowalski \shortcite{Kow79} advocates the separation of the
{\em logic} and {\em control} aspects of a logic program and has
coined the famous formula
\begin{quote}
Algorithm = Logic + Control.
\end{quote}
The programmer should be responsible for the logic part, and hence a
logic program should be a (first order logic) specification. The
control should be taken care of by the logic programming system.  One
aspect of control in logic programs is the {\em selection rule}. This
is a rule stating which atom in a query is selected in each derivation
step.  It is well-known that soundness and completeness of
SLD-resolution is independent of the selection rule \cite{Apt97}.  
However, a stronger property is usually required
for a selection rule to be useful in programming, namely termination.

\begin{definition} \label{c2:def:comp}
 A {\em terminating control\/} for a program $P$ and a query $Q$ is a
 selection rule $s$ such that every SLD-derivation of $P$ and $Q$ via
 $s$ is finite.  
\end{definition}

 In reality, logic programming is far from the ideal that the logic
and control aspects are separated. Without the programmer being aware
of the control and writing programs accordingly, logic programs would
usually be hopelessly inefficient or even non-terminating.

The usual selection rule of early systems is the {\em LD} selection
rule: in each derivation step, the leftmost atom in a query is
selected for resolution. This selection rule is based on the
assumption that programs are written in such a way that the data flow
within a query or clause body is from left to right. Under this
assumption, this selection rule is usually a terminating control. For
most applications, this selection rule is appropriate in that it
allows for an efficient implementation.

Second generation logic languages adopt more flexible control
primitives, which allow for addressing logic and control separately.
Program clauses have their usual logical reading.
In addition, programs are augmented by {\em delay declarations\/} or
{\em annotations\/} that specify restrictions on the admissible
selection rules. These languages include NU-Prolog \cite{nuprolog},
G{\"o}del \cite{HL94} \index{G{\"o}del} and Mercury \cite{SHC96}.

In this survey, we classify programs and queries according to the
selection rules under which they terminate, hence investigating the
influence of the selection rule on termination. As most approaches to
the termination problem, we are interested in {\em universal}
termination of logic programs and queries, that is, showing that {\em
  all} derivations for a program and query (via a certain selection
rule) are finite. This is in contrast to {\em existential} termination
\cite{Bau92,SD94,Mar96}.  Also, we consider \emph{definite} logic
programs, as opposed to logic programs that also contain negated
literals in clause bodies.

Figure~\ref{overview-fig} gives an overview of the classes we
consider. Arrows drawn with solid lines stand for set inclusion 
(``$\to$ corresponds to $\subset$''). 
The numbers in the figure correspond to propositions in
section~\ref{relations-sec}. 

\setlength{\unitlength}{0.35cm}
\begin{figure}\figrule
\begin{center}
\begin{picture}(33,30)
\put(16,1){\vector(0,1){0}}
\put(8,1){\makebox(16,1){\em (Strong Termination)}}
\put(8,2){\makebox(16,1){\bf Recurrent Programs}}
\put(14,3){\vector(-1,1){8}}
\put(16,3){\vector(0,1){13}}
\put(14,5){\makebox(2,1){\ref{c2:r:rec}}}
\put(18,3){\vector(1,1){3}}
\put(16,6){\makebox(16,1){\em (Input Termination)}}
\put(16,7){\makebox(16,1){\bf Simply-Acceptable Programs}}
\dashline{0.3}(20,8)(12,11)
\put(12,11){\vector(-3,1){0}}
\put(18,8.7){\makebox(2,1){\ref{input-implies-local-prop}}}
\dashline{0.3}(24,8)(18,16)
\put(18,16){\vector(-3,4){0}}
\put(21,12){\makebox(2,1){\ref{input-implies-left-prop}}}
\dashline{0.3}(28,8)(20,21)
\put(20,21){\vector(-2,3){0}}
\put(25,13){\makebox(2,1){\ref{input-implies-exists-prop}}}
\put(0,11){\makebox(16,1){\em (Local delay termination)}}
\put(0,12){\makebox(16,1){\bf Delay-Recurrent Programs}}
\dashline{0.3}(8,13)(14,16)
\put(14,16){\vector(2,1){0}}
\put(8,14){\makebox(2,1){\ref{local-implies-left-prop}}}
\dashline{0.3}(4,13)(12,21)
\put(12,21){\vector(1,1){0}}
\put(6,17){\makebox(2,1){\ref{local-implies-exists-prop}}}
\put(8,16){\makebox(16,1){\em (Left-Termination)}}
\put(8,17){\makebox(16,1){\bf Acceptable Programs}}
\put(16,18){\vector(0,1){3}}
\put(14,19){\makebox(2,1){\ref{c2:r:rec}}}
\put(8,21){\makebox(16,1){\em (Exists-Termination)}}
\put(8,22){\makebox(16,1){\bf Fair-Bounded Programs}}
\put(16,23){\vector(0,1){3}}
\put(14,24){\makebox(2,1){\ref{c2:r:rec}}}
\put(8,26){\makebox(16,1){\em (Bounded Nondeterminism)}}
\put(8,27){\makebox(16,1){\bf Bounded Programs}}
\dottedline{0.4}(16,28)(16,29)(33,29)(33,0)(16,0)(16,1)
\put(31,24){\makebox(2,1){\ref{c2:bn:transfcorr}}}
\put(19,29){\makebox(14,1){\bf Inference+Transformation}}
\end{picture}
\end{center}
\caption{An overview of the classes\label{overview-fig}}
\figrule\end{figure}
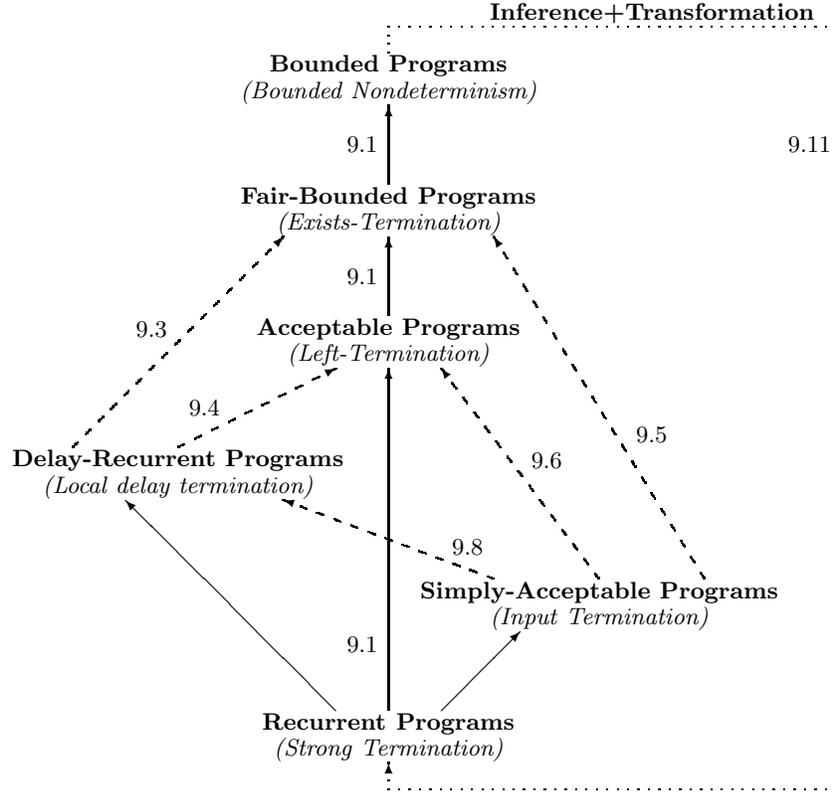

A program $P$ and query $Q$ {\em strongly terminate} if they terminate
for {\em all} selection rules. This class of programs has been studied
mainly by Bezem \shortcite{Bez93}.  Naturally, this class is the
smallest we consider.  A program $P$ and query $Q$ {\em
left-terminate} if they terminate for the {\em LD} selection rule.
The vast majority of the literature is concerned with this class;
see \cite{SD94} for an overview.  A program $P$ and query $Q$ {\em
$\exists$-terminate} if there {\em exists} a selection rule for which
they terminate. This notion of termination has been introduced by
Ruggieri \shortcite{RugTCS98,Rug99}.  Surprisingly, this is still not the
largest class we consider. Namely, there is the class of programs for
which there are only finitely many {\em successful} derivations
(although there could also be infinite derivations). We say that these
programs have {\em bounded nondeterminism}, a notion studied by
Pedreschi \& Ruggieri \shortcite{PR99}. Such programs can be transformed
into equivalent programs which strongly terminate, as indicated in the
figure and stated in Theorem~\ref{c2:bn:transfcorr}.

To explain the two remaining classes shown in the figure, and their
relationship with left-terminating programs, we have to introduce the
concept of {\em modes}. A mode is a labelling of each argument
position of a predicate as either input or output. It indicates the
intended data flow in a query or clause body.

An {\em input-consuming} derivation is a derivation where an atom can
be selected only when its input arguments are instantiated to a
sufficient degree, so that unification with the head of the clause
does not instantiate them further. A program and a query {\em
input terminate} if all input-consuming
derivations for this program and query are finite. This class of
programs has been studied by Smaus \shortcite{Sma99} and Bossi 
{\em et al.} \shortcite{BER99,BER00,BERS01}.

A {\em local} selection rule is a selection rule specifying that an
atom can only be selected if there is no other atom which was
introduced (by resolution) more recently.  Marchiori \&
Teusink~\shortcite{MT99} have
studied termination for selection rules that are both local and 
{\em delay-safe}, i.e., they respect the {\em delay
declarations}. We will call termination w.r.t.~such
selection rules {\em local delay} termination.  A priori, the LD
selection rule, input-consuming selection rules and local delay-safe
selection rules are not formally comparable.  Under reasonable
assumptions however, one can say that assuming input-consuming
selection rules is weaker than assuming local and delay-safe selection
rules, which is again weaker than assuming the LD selection rule. This
is indicated in the figure by arrows drawn with dashed lines.
Again, the numbers in the figure correspond to propositions in
section~\ref{relations-sec}. 

In this survey, we present declarative characterisations of the
classes of programs and queries that terminate with respect to each of
the mentioned notions of termination. The characterisations make use
of level mappings and Herbrand models in order to provide proof
obligations on program clauses and queries. All characterisations are
sound. Except for the case of local delay termination, they are also
complete (in the case of input termination, this holds only under
certain restrictions).

This survey is organised as follows. The next section introduces some
basic concepts and fixes the notation. Then we have six sections
corresponding to the six classes in figure~\ref{overview-fig}, defined
by increasingly strong assumptions about the selection rule.  In each
section, we introduce a notion of termination and provide a
declarative characterisation for the corresponding class of
terminating programs and queries. In section~\ref{relations-sec}, we
establish relations between the classes, formally showing the
implications of figure~\ref{overview-fig}.  Section~\ref{related-sec}
discusses the related work, and section~\ref{conc-sec} concludes.

\section{Background and Notation}\label{prelim-sec}

We use the notation of Apt \shortcite{Apt97}, when not otherwise
specified. In particular, throughout this article we consider a fixed
language $L$ in which programs and queries are written. All the
results are {\it parametric\/} with respect to $L$, provided that $L$
is rich enough to contain the symbols of the programs and queries
under consideration.  We denote with $U_L$ and $B_L$ the Herbrand
universe and the Herbrand base on $L$. $Term_L$ and $Atom_L$ denote
the set of terms and atoms on $L$. 
We use typewriter font for logical variables,
e.g.~$\mathtt{X},\mathtt{Ys}$, 
upper case letters for arbitrary terms, e.g.  $\mathit{Xs}$, and 
lower case letters for ground terms, e.g.~$t, x, \mathit{xs}$.
We denote by $\inst_L(P)$ ($\groun{L}{P}$) the set of (ground)
instances of all clauses
in $P$ that are in language $L$. The notation
$\groun{L}{Q}$ for a query $Q$ is defined analogously. 

The domain (resp., set of variables in the range) 
of a substitution $\theta$ is denoted as $\dom(\theta)$ (resp., 
$\codom(\theta)$). 

\subsection{Modes}\label{modes-subsec}
For a predicate $p/n$, a {\em mode} is an atom $p(m_1,\dots,m_n)$,
where $m_i \in \{{\tt \inp,\out} \}$ for $i \in [1,n]$.  Positions
with $\inp$ are called {\em input positions}, and positions with
$\out$ are called {\em output positions} of $p$.  To simplify the
notation, an atom written as $p(\vect{s},\vect{t})$ means: $\vect{s}$
is the vector of terms filling in the input positions, and $\vect{t}$ is
the vector of terms filling in the output positions.  An atom
$p(\vect{s},\vect{t})$ is {\em input-linear} if $\vect{s}$ is linear,
i.e.~each variable occurs at most once in $\vect{s}$. The atom is 
{\em output-linear} if $\vect{t}$ is linear.

In the literature, several correctness criteria concerning the modes
have been proposed, the most important ones being nicely-modedness and 
well-modedness~\cite{Apt97}. In this article, we need {\em simply}
moded programs~\cite{AE93}, which are a special case of nicely 
moded programs, as well as {\em well moded} programs.

\begin{definition}
\label{simply-moded-def}
A clause 
$p(\vect{t}_0,\vect{s}_{n+1})\la 
p_1(\vect{s}_1,\vect{t}_1),\ldots,p_n(\vect{s}_n,\vect{t}_n)$
is 
  \emph{simply moded} if $\vect{t}_1,\ldots,\vect{t}_n$ is a linear
  vector of variables and for all $i\in [1,n]$
 
  $${\it Var}(\vect{t}_i)\cap {\it Var}(\vect{t}_0) = \emptyset
\quad \mbox{and} \quad
{\it Var}(\vect{t}_i)\cap 
\bigcup_{j=1}^i{\it Var}(\vect{s}_j)=\emptyset.$$
  
A query $\vect{B}$ is  \emph{simply moded} if the clause $q
  \la \vect{B}$ is simply moded, where $q$ is any variable-free atom.
A program is simply moded if all of its clauses are.

A query (clause, program) is {\em permutation simply moded} if it is
simply moded modulo reordering of the atoms of the query (each clause body).
\end{definition}

Thus, a clause is simply moded if the output positions of body atoms
are filled in by distinct variables, and every variable occurring in
an output position of a body atom does not occur in an earlier input
position.  In particular, every unit clause is simply moded.

\begin{definition}\label{well-moded}
A query $Q =
p_1(\vect{s}_1,\vect{t}_1),\dots,p_n(\vect{s}_n,\vect{t}_n)$ is {\em
well moded} if for all $i \in [1,n]$ and $K=1$
\begin{equation}
\vars(\vect{s}_i) \subseteq
\bigcup_{j = K}^{i-1}
  \vars(\vect{t}_j) \label{wm-eq}
\end{equation}
The clause 
$p(\vect{t}_0,\vect{s}_{n+1}) \gets Q$
is {\em well moded} if (\ref{wm-eq}) holds
for all $i \in [1,n+1]$ and $K = 0$.
A program is {\em well moded} if all of its clauses are well moded.

A query (clause, program) is {\em permutation well moded} if it is
well moded modulo reordering of the atoms of the query (each clause body).
\end{definition}

Almost all programs we consider in this article are permutation well
and 
simply moded with respect to the same set of modes.  The program in
figure~\ref{transp-prog} is an exception due to the fact that our
notion of modes cannot capture that sub-arguments of a term can have
different modes. We do not always give the modes explicitly, but they are
usually easy to guess.


\subsection{Selection Rules}
Let $\mathit{INIT}$ be the set of initial fragments of SLD-derivations in which
the last query is non-empty. The standard definition of 
{\em selection rule\/} is as follows: a selection rule is a function 
that, when applied to an element in $\mathit{INIT}$, yields an
occurrence of an atom in its last query \cite{Apt97}. In this article, 
we assume an extended definition: 
we also allow that a selection rule may select no
atom (a situation called {\em deadlock}), and we allow that it
not only returns the selected atom, but also specifies the set of program
clauses that may be used to resolve the atom. Whenever we want to
emphasise that a selection rule always selects exactly one atom together
with the entire set of clauses for that atom's predicate, we speak of a 
{\em standard selection rule}.
Note that for the extended definition, completeness of SLD-resolution is lost in general.
Selection rules are denoted by $s$.

We now define the selection rules used in this article, except for
{\em delay-safe} selection rules, since these
rely on notions introduced only later.

Input-consuming selection rules are defined w.r.t.~a given mode.  A
selection rule $s$ is {\em input-consuming} for a program $P$ if 
either 
\begin{itemize}
\item
$s$ selects an atom $p(\vect{s},\vect{t})$ and a non-empty set of
clauses of $P$ such that $p(\vect{s},\vect{t})$
and each head of a clause in the set are unifiable with an mgu
$\sigma$, and $\dom(\sigma) \cap \vars(\vect{s}) = \emptyset$, or
\item
$s$ selects an atom $p(\vect{s},\vect{t})$ that unifies with no clause head from $P$,
together with all clauses in $P$ (this models {\em failure}), or
\item
if the previous cases are impossible, $s$ selects no atom 
(i.e.\ we have {\em deadlock}).
\end{itemize}
Consider a query, containing atoms $A$ and $B$, in an initial fragment
$\xi$ of a derivation. Then $A$ is {\em introduced more recently} than
$B$ if the derivation step introducing $A$ comes after the step
introducing $B$, in $\xi$.  A {\em local selection rule} is a
selection rule that specifies that an atom in a query can be selected
only if there is no more recently introduced atom in the query.

The usual {\em LD} selection rule (also called {\em leftmost}
selection rule) always selects the leftmost atom in the last query of
an element in $\mathit{INIT}$. The {\em RD} selection rule (also
called {\em rightmost}) always selects the rightmost atom.

A standard selection rule $s$ is {\em fair\/} if for every
SLD-derivation $\xi$ via $s$ either $\xi$ is finite or for every atom
$A$ in $\xi$, (some further instantiated version of) $A$ is eventually
selected.

\subsection{Universal Termination}
In general terms, the problem of universal termination of a
program $P$ and a query $Q$ w.r.t.~a set of admissible selection rules
consists of showing that every rule in the set is a terminating
control for $P$ and $Q$.

\begin{definition}\label{c2:def:ut}
  A program $P$ and a query $Q$ {\em universally terminate\/} w.r.t.~a
  set of selection rules ${\cal S}$ if every SLD-derivation of $P$ and
  $Q$ via any selection rule from ${\cal S}$ is finite. 
\end{definition}

 Note that, since SLD-trees are finitely branching, by K{\"o}nig's
Lemma, ``every SLD-derivation for $P$ and $Q$ via a selection rule $s$
is finite'' is equivalent to stating that the SLD-tree of $P$ and $Q$
via $s$ is finite.

We say that a class of programs and queries is a {\em sound}
characterisation of universal termination w.r.t.~${\cal S}$ if every
program and query in the class universally terminate w.r.t.~${\cal
S}$. Conversely, it is {\em complete} if every program and query that
universally terminate w.r.t.~${\cal S}$ are in the class.

\subsection{Norms and Level Mappings}\label{norms-subsec}
All the characterisations of terminating programs we propose make
use of the notions of norm and level mapping 
\cite{Cav89}. Depending on the
approach, such notions are defined on ground or arbitrary objects.

In the following definition, $Term_L/\!\!\sim$ 
denotes the set of equivalence classes of terms modulo variance.
Similarly, we define $Atom_L/\!\!\sim$. 

\begin{definition} \label{c2:def:nlm}
A {\em norm\/} is a function $|.|: U_L \ra \nat$.  A 
{\em level mapping\/} is a function $|.|: B_L \ra \nat$.  For a ground
atom $A$, $|A|$ is called the level of $A$.  

An atom $A$ is {\em bounded} w.r.t.~the level mapping 
$|.|$ if there exists $k \in \nat$ such that for every 
$A' \in \groun{L}{A}$, we have $k > |A'|$.  

A {\em generalised norm\/} is a function 
$|.|: \mathit{Term}_L/\!\!\sim \ra \nat$. A 
{\em generalised level mapping\/} is a function 
$|.|: \mathit{Atom}_L/\!\!\sim \ra \nat$ . Abusing notation, 
we write $|T|$ ($|A|$) to denote the value
of $|.|$ on the equivalence class of the term $T$ (the atom $A$).
\end{definition}


(Generalised) level mappings are used to measure the ``size'' of a
query and show that this size decreases along a derivation, hence
showing termination. They are usually defined based on (generalised)
norms. 
Therefore we often use the same notation $|.|$ for a norm and a level
mapping based on it.

Of course, a generalised norm or level mapping can be interpreted as
an ordinary norm or level mapping by restricting its domain to ground
objects. Therefore, we now give some examples of {\em generalised}
norms and level mappings.  

One commonly used generalised norm is the term size norm, defined as
\[
\begin{array}{rcll}
size( f(\tn{T}{n}) )            & = &  1 + size(T_1) + \ldots + size(T_n) &
\mbox{if $n > 0$}\\
size(T)                         & = & 0 &
\mbox{if $T$ constant/variable.}
\end{array}
\]
Intuitively, the size of a term $T$ is the number of
function symbols occurring in $T$, excluding constants.
Another widely used norm is the list-length function, defined as 
\[
\begin{array}{rcll}
 | [T|Ts] |     & = & 1 + |Ts| \\
 | f(\ldots) | & = & 0 & \mbox{if}\ f \neq  [\: .\: |\: .\: ].
\end{array}
\]
In particular, for a nil-terminated list $[T_1,\ldots,T_n]$, the
list-length is $n$.  

We will see later that usually, level mappings measure the {\em input} 
arguments of a query, even though this is often just an intuitive
understanding and not explicit. Moreover, the choice of a particular
selection rule often reflects a particular mode of the program. 
In this sense, the choice of the level mapping must depend on the  
selection rule, via the modes. This will be seen in our examples. 
 
However, apart form the dependency just mentioned, the choice of level
mapping is an aspect of termination which is rather independent from
the choice of the selection rule. In particular, one does not find any
interesting relationship between the underlying {\em norms} and the
selection rule. This is why the detailed study of various norms and
level mappings is beyond the scope of this article, although it is an important
aspect of automated proofs of termination \cite{DDF93,BCF94}.

We now define level mappings where the dependency on the modes is made 
explicit \cite{EBC99}.

\begin{definition} \label{moded-level-mapping}
A {\em moded (generalised) level mapping} 
$|.|$ is a (generalised) level
mapping such that for any (not necessarily) ground
$\vect{s}$, $\vect{t}$ and $\vect{u}$, 
$|p(\vect{s},\vect{t})| = |p(\vect{s},\vect{u})|$.
\end{definition}

The condition 
$|p(\mathbf{s},\mathbf{t})|=|p(\mathbf{s},\mathbf{u})|$ states 
that the \emph{level} of an atom is independent from
the terms in its output positions. 

\subsection{Models}
Several of the criteria for termination we consider rely on
information supplied by a model of the program under consideration.
We provide the definition of Herbrand interpretations and models
\cite{Apt97}.

A {\em Herbrand interpretation} $I$ is a set of ground atoms. A ground
atom $A$ is {\em true in $I$}, written $I\models A$, if $A\in I$. This
notation is extended to ground queries in the obvious way.  $I$ is a
Herbrand {\em model} of program $P$ if for each $\clausab \in
\groun{L}{P}$, we have that $I\models B_1,\dots,B_n$ implies $I\models
A$.  

When speaking of the {\em least} Herbrand model of $P$, we mean 
least w.r.t.~set inclusion.  In termination analysis, it is usually not
necessary to consider the least Herbrand model, which may be difficult
or impossible to determine.  Instead, one uses
models that capture some {\em argument size relationship} between the
arguments of each predicate \cite{SD94}. For example, a model for the
usual $\mathtt{append}$ predicate is
$$\{\mathtt{append}(xs,ys,zs) \mid  |zs| = |xs| + |ys|  \}$$
where $|.|$ is the list-length function.

\section{Strong Termination}\label{strong-sec}

\termination
Early approaches to the termination problem treated universal
termination w.r.t.\ {\em all} selection rules, called {\em strong}
termination. Generally speaking, strongly terminating programs and
queries are either very trivial or especially written for theoretical
considerations.

\begin{definition} \label{c2:def:strongt}
  A program $P$ and query $Q$ {\em strongly terminate} if they
  universally terminate w.r.t.~the set of all selection rules.  
\end{definition}

\decrease
 In the following, we recall the approach of Bezem
\shortcite{Bez93}, who defined the class of
recurrent programs and queries.  Intuitively, a program is recurrent
if for every ground instance of a clause, the level of the body atoms
is smaller than the level of the head.

\begin{definition} \label{c2:def:recp}\label{c2:def:recq}
  Let $|.|$ be a level mapping.  

A program $P$ is {\em recurrent by}
  $|.|$ if for every $\clausab$ in $\groun{L}{P}$:
\[
\mbox{ for $i \in [1,n]$} \quad |A| > |B_i|.
\]
A query $Q$ is {\em recurrent by $|.|$\/} if there exists 
$k \in \nat$ such that for
  every  $\tn{A}{n} \in \groun{L}{Q}$:
\[  
\mbox{ for $i \in [1,n]$ } \quad k > |A_i|. 
\]
\end{definition}

In the above definition, the proof obligations for a query $Q$ are
derived from those for the program $\{ {\tt p} \la Q \}$, where {\tt
  p} is a fresh predicate symbol. Intuitively, this is justified by
the fact that the termination behaviour of the query $Q$ and a program
$P$ is the same as for the query {\tt p} and the program $P \U \{ {\tt
  p} \la Q\}$.  So $k$ plays the role of the level of the atom {\tt
  p}. In the original work \cite{Bez93}, the query was called
{\em bounded}. Throughout the
paper, we prefer to maintain a uniform naming convention both for
programs and queries.

In subsection~\ref{characterisations-subsec}, we will compare
recurrence to other characterisations.

\maintheorem
Termination properties of recurrent programs are summarised in
the following theorem.

\begin{theorem}[\cite{Bez93}] \label{c2:rec:thm}
  Let $P$ be a  program and $Q$ a query.

  If $P$ and $Q$ are both recurrent by a  level mapping $|.|$,
  then they strongly terminate.

  Conversely, if $P$ and every {\em ground query\/} strongly
  terminate, then $P$ is recurrent by some level mapping $|.|$.  If
  in addition $P$ and $Q$ strongly terminate, then $P$ and $Q$ are
  both recurrent by some level mapping $|.|$.  
\end{theorem}
\begin{proof}
The result is shown in \cite{Bez93} for standard selection
rules. It easily extends to our generalisation of selection rules by
noting that $P$ and $Q$ strongly terminate iff they universally
terminate w.r.t.\ the set of standard selection rules. The only-if
part is immediate. The if-part follows by noting that a derivation
via an arbitrary selection rule is a (prefix of a) derivation via a 
{\em standard} selection rule.  
\end{proof}

\examples
\begin{example} \label{ex:sat-is-recurrent}
The program {\tt SAT} in figure~\ref{sat-prog} 
decides propositional satisfiability.
The program is readily checked to be recurrent by $|.|$, where
we define
 \[ 
|\mathtt{sat}(t)| = |\mathtt{inval}(t)| =
 size(t). 
\]
Note that Definition~\ref{c2:def:recp} imposes no proof obligations for unit
clauses. The query {\tt sat($X$)} is recurrent iff there exists a natural
$k$ such that for every ground instance
$x$ of $X$, we have that $size(x)$ is bounded by $k$. 
Obviously, this is the case iff $X$ is already a ground term. For
instance, the query {\tt sat(not(true) $\land$ false)} is recurrent,
while the query {\tt sat(false $\land$ X)} is not.  
\end{example}

\begin{figure}\figrule
\begin{minipage}[t]{5cm}
\begin{janprogram}
\%\> sat(Formula) $\la$ \\
\%\> \>  \mbox{\rm there is a true instance of} Formula\\[2ex]
\> sat(true).\\
\> sat(X $\land$ Y) \la \\
\> \> sat(X), sat(Y).\\
\> sat(not X) \la \ inval(X).
\end{janprogram}
\end{minipage}
\hfill
\begin{minipage}[t]{5cm}
\begin{janprogram}
\> inval(false).\\
\> inval(X $\land$ Y) \la \ inval(X).\\
\> inval(X $\land$ Y) \la \ inval(Y).\\
\> inval(not X) \la \  sat(X).
\end{janprogram}
\end{minipage}
\caption{\tt SAT\label{sat-prog}}
\figrule\end{figure}

 Note that the choice of an appropriate level
mapping depends on the intended mode of the program and query. 
Even though this is usually not explicit, level mappings measure 
the size of the {\em input} arguments of an atom \cite{EBC99}.

\begin{example} \label{ex:append-is-recurrent}
Figure~\ref{append-prog} shows the {\tt APPEND} program.
It is easy to check that {\tt APPEND} is recurrent by the level
mapping $|\mathtt{append}(xs,ys,zs)| = |xs|$ and also by
$|\mathtt{append}(xs,ys,zs)| = |zs|$ 
(recall that 
$|.|$ is the list-length function).  A query 
{\tt append($\mathit{Xs}$,$\mathit{Ys}$,$\mathit{Zs}$)} is recurrent 
by the first level mapping iff
$\mathit{Xs}$ is a list, and by the second iff $\mathit{Zs}$ is a list.  The level
mapping
\[ 
|\mathtt{append}(xs,ys,zs)| = min\{ |xs|, |zs| \} 
\]
combines the advantages of both level mappings.  {\tt APPEND} is
easily seen to be recurrent by it, and if $\mathit{Xs}$ 
{\em or\/} $\mathit{Zs}$
is a list, 
$\mathtt{append}(\mathit{Xs},\mathit{Ys},\mathit{Zs})$ is recurrent by it.  
\end{example}

\begin{figure}\figrule
\begin{janprogram}
\%\>  append(Xs,Ys,Zs) \la \ \\
\%\> \> {\rm {\tt Zs} is the result of concatenating the
lists {\tt Xs} and {\tt Ys}.} \\[2mm]
\>  append([],Ys,Ys). \\
\>  append([X|Xs],Ys,[X|Zs]) \la \ append(Xs,Ys,Zs). 
\end{janprogram}
\caption{\tt APPEND\label{append-prog}}
\figrule\end{figure}

\incompleteness \label{recurrency-restricted-completeness}
 Note that completeness is not stated in full general terms,
i.e.~recurrence is not a complete proof method for strong termination.
 Informally speaking, incompleteness is due to the use of level
mappings, which are functions that must specify a value for every
ground atom. 
Therefore, if $P$ strongly terminates for a certain ground query $Q$
but not for all ground queries, we cannot conclude that $P$ is recurrent.
We provide a general
completeness result in section~\ref{left-sec} for a class of
programs containing recurrent programs.

\section{Input Termination}\label{IC-sec}
\limitation
We have said above that the class of strongly terminating programs and
queries is very limited. Even if a program is recurrent, it may not
strongly terminate for a query of interest since the query is not
recurrent.

\begin{example} \label{ex:even-query-is-not-recurrent}
The program {\tt EVEN} in figure~\ref{even-prog} 
is recurrent by defining
\begin{eqnarray*}
|\mathtt{even}(x)|   & = & size(x)\\[-0.5ex]
|\mathtt{lte}(x,y) | & = & size(y).
\end{eqnarray*}
Now consider the query 
$Q = \mathtt{even(X),\ lte(X,s^\mathit{100}(0))}$,
which is supposed to compute the even numbers not exceeding 100.  By
always selecting the leftmost atom, one can easily obtain an infinite
derivation for {\tt EVEN} and $Q$.  As a consequence of
Theorem~\ref{c2:rec:thm}, $Q$ is not recurrent.  
\end{example}

\begin{figure}\figrule
\begin{minipage}[t]{5cm}
\begin{janprogram}
\%\>  even(X) \la \ \\
\%\> \> {\tt X} {\rm is an even natural number.} \\[2mm]
\>  even(s(s(X))) \la \ even(X). \\
\>  even(0).
\end{janprogram}
\end{minipage}
\hfill
\begin{minipage}[t]{5cm}
\begin{janprogram}
\%\>  lte(X,Y) \la \ \\
\%\> \>  X,Y {\rm are natural numbers}\\
\%\> \> {\rm s.t. {\tt X} is
smaller or equal 
than {\tt Y}}.\\[2mm]
\>  lte(s(X),s(Y)) \la \ lte(X,Y). \\
\>  lte(0,Y).
\end{janprogram}
\end{minipage}
\caption{\tt EVEN\label{even-prog}}
\figrule\end{figure}

\termination
We now define termination for input-consuming
derivations~\cite{BERS01}, i.e.~derivations via an input-consuming
selection rule.

\begin{definition} \label{c2:def:ict}
  A program $P$ and query $Q$ {\em input terminate\/} if
  they universally terminate w.r.t.~the set consisting of 
the input-consuming selection rules.  
\end{definition}

The requirement of input-consuming derivations merely
reflects the very meaning of {\em input}: an atom must only consume
its own input, not produce it. In existing implementations,
input-consuming derivations can be ensured using control constructs
such as delay-declarations \cite{HL94,sicstus,SHC96,nuprolog}.

In the above example, the obvious mode is 
$\tt even(\inp),\ lte(\out,\inp)$.  
With this mode, we will show that {\tt EVEN} and
$Q$ input terminate.  If we assume a selection rule
that is input-consuming while always selecting the leftmost atom if
possible, then the above example is a contrived instance of the {\em
  generate-and-test} paradigm. This paradigm involves two procedures,
one which generates a set of candidates, and another which tests
whether these candidates are solutions to the problem. The test occurs
to the left of the generator so that tests take place as soon as
possible, i.e.~as soon as sufficient input has been generated for the 
derivation to be input-consuming.

Proofs of input termination differ from proofs of strong
termination in an important respect. For the latter, we require that
the initial query is recurrent, and as a consequence we have that all
queries in any derivation from it are recurrent (we say that
recurrence is {\em persistent} under resolution). This means that, at
the time an atom is selected, the depth of its SLD tree is bounded. In
contrast, input termination does not need such a strong
requirement on each selected atom.

\begin{example}\label{non-pindown-ex}
  Consider the {\tt EVEN} program and the following input-consuming
  derivation, where we underline the selected atom in each step
\[ 
\begin{array}{l}
\tt even(X),\ \underline{lte(X,s^\mathit{100}(0))} \longrightarrow \quad
even(s(X')),\ \underline{lte(X',s^\mathit{99}(0))} \longrightarrow\\
\tt \underline{even(s(s(X'')))},\ lte(X'',s^\mathit{98}(0)) \longrightarrow \quad
even(X''),\ \underline{lte(X'',s^\mathit{98}(0))} \dots
\end{array}
\] 
At the time when $\tt even(s(s(X'')))$ is selected, the depth of its
SLD-tree is not bounded (without knowing the eventual instantiation of
$\tt X''$).  
\end{example}

\onlyhere{Information on Data Flow: Simply-local Substitutions and Models}
 Since the depth of the SLD-tree of the selected atom depends on
further instantiation of the atom, it is important that programs are
well-behaved w.r.t.~the modes.  This is illustrated in the following
example.

\begin{example} \label{circular-ex}
Consider the {\tt APPEND} program 
in mode $\tt append(\inp,\inp,\out)$ and the query 
\[
\tt append([1|As],[],Bs), append(Bs,[],As).
\]
Then we have the following infinite 
input-consuming derivation:
\[
\begin{array}{l}
\tt \underline{append([1|As],[],Bs)},\ append(Bs,[],As) \longrightarrow \\
\tt append(As,[],Bs'),\ \underline{append([1|Bs'],[],As)} \longrightarrow \\
\tt \underline{append([1|As'],[],Bs')},\ append(Bs',[],As') \longrightarrow 
\dots
\end{array}
\]
This well-known termination problem of programs with coroutining has
been identified as {\em circular modes} by Naish \shortcite{Nai92c}. 
\end{example}

 To avoid the above situation, we require programs to be simply moded
 (see subsection~\ref{modes-subsec}).

We now define \emph{simply-local} substitutions,
which reflect the way simply moded clauses become instantiated in
input-consuming derivations. Given a clause 
$\cequals p(\vect{t}_0,\vect{s}_{n+1})\la 
p_1(\vect{s}_1,\vect{t}_1),\ldots,p_n(\vect{s}_n,\vect{t}_n)$ used
in an input-consuming derivation, first $\vect{t}_0$ becomes
instantiated, and the range of that substitution contains only
variables from outside of $c$. Then, by resolving  
$p_1(\vect{s}_1,\vect{t}_1)$, the vector $\vect{t}_1$ becomes
instantiated, and the range of that substitution contains variables
from outside of $c$ in addition to variables from $\vect{s}_1$.
Continuing in the same way, finally,  by resolving  
$p_n(\vect{s}_n,\vect{t}_n)$, the vector $\vect{t}_n$ becomes
instantiated, and the range of that substitution contains variables
from outside of $c$ in addition to variables from $\vect{s}_1\dots\vect{s}_n$.
A substitution is \emph{simply-local} if it is composed from
substitutions as sketched above. The formal definition is as follows.

\begin{definition}
\label{def:simply-local}
A substitution $\theta$ is \emph{simply-local} w.r.t.\ 
the clause 
$\cequals p(\vect{t}_0,\vect{s}_{n+1})\la 
p_1(\vect{s}_1,\vect{t}_1),$ $\ldots,$ $p_n(\vect{s}_n,\vect{t}_n)$
 if 
there exist substitutions $\sigma_0,\sigma_1\ldots,\sigma_n$
and disjoint sets of fresh (w.r.t.\  $c$) variables
$v_0,v_1,\ldots,v_n$   
such that
 $\theta=\sigma_0\sigma_1\cdots\sigma_n$ 
where for $i\in \{0,\ldots,n\}$,
\begin{itemize}
\item $\dom(\sigma_i)\subseteq \vars(\vect{t}_i)$,
\item  $\codom(\sigma_i)\subseteq  
\vars(\vect{s}_i \sigma_0\sigma_1\cdots\sigma_{i-1}) \cup
v_i$.\footnote{
Note that $\vect{s}_0$ is undefined. By abuse of notation, 
$\vars(\vect{s}_0 \dots) = \emptyset$.}
\end{itemize}

$\theta$ is \emph{simply-local} w.r.t.\ a query $\vect{B}$ if
  $\theta$ is simply-local w.r.t.\ the clause $q \la \vect{B}$ where $q$ is
  any variable-free atom. 
\end{definition}

Note that in the case of a simply-local substitution w.r.t.~a query, $\sigma_0$ is
the empty substitution, since 
$\mathit{Dom}(\sigma_0)\subseteq \mathit{Var}(q)$ where $q$ is an (imaginary)
variable-free atom.  Note also that if  
$\vect{A},B,\vect{C}
\longrightarrow
(\vect{A},\vect{B},{\bf C})\theta$ 
is an input-consuming derivation step  using clause
$\cequals H\gets \vect{B}$, then 
$\theta_{|H}$ is simply-local w.r.t.~the clause $H\la$ and 
$\theta_{|B}$ is simply-local w.r.t.~the atom $B$~\cite{BERS01}.

\begin{example}\label{ex:local-substitutions}
Consider \texttt{APPEND} in mode $\mathtt{append}(\inp,\inp,\out)$, and its
recursive clause 
\[
\cequals \tt append([H|Xs],Ys,[H|Zs]) \la append(Xs,Ys,Zs).
\]
The substitution 
$\theta = \{\tt H/V, Xs/[], Ys/[W], Zs/[W]\}$ is simply-local w.r.t.\ $c$: let
$\sigma_0 = \{\tt H/V, Xs/[], Ys/[W]\}$ and
$\sigma_1 = \{\tt Zs/[W]\}$; then
$\dom(\sigma_0)\subseteq \{\tt H, Xs, Ys\}$, and
$\codom(\sigma_0)\subseteq v_0$ where 
$v_0 = \{\tt V, W\}$, and
$\dom(\sigma_1)\subseteq \{\tt Zs\}$, and
$\codom(\sigma_1)\subseteq 
  \vars({\tt (Xs,Ys)}\sigma_0)$.
\end{example}

Based on simply-local substitutions, we now define a restricted notion
of model.

\begin{definition}\label{sl-model-def}
Let $I\subseteq \mathit{Atom}_L$.
We say that $I$ is a \emph{simply-local model} of 
$\cequals H\gets B_1,\ldots,B_n$ if for every substitution 
$\theta$ simply-local w.r.t.\ $c$, 
\begin{equation}
\mbox{if $B_1\theta,\ldots,B_n\theta\in I$ then $ H\theta\in I$.}
\label{SL-model-eq}
\end{equation}
$I$ is a \emph{simply-local model} of a program $P$ if it is a
  simply-local model of each clause of it.  
\end{definition}

Note that a simply-local model 
is not necessarily a model in the classical sense, since 
$I$ is not necessarily a set of ground atoms, and  
the substitution in (\ref{SL-model-eq}) is required to be 
simply-local. 
For example, given the program
$\{\tt q(1),\; p(X) \!\la\! q(X)\}$
with modes 
$\tt q(\inp),\ p(\out)$, a model must contain the atom
$\tt p(1)$, whereas a simply-local model does not necessarily contain
$\tt p(1)$, since $\{\tt X/1\}$ is not simply-local w.r.t.\ 
$\tt p(X) \la q(X)$.
The next subsection will further clarify the role of simply-local
models.

Let $\mathit{SM}_P$ be the set of all simply moded atoms in 
$\mathit{Atom}_L$.
It has been shown that the 
least simply-local model of $P$
containing $\mathit{SM}_P$ exists and can be computed by a variant of
the well-known $T_P$-operator~\cite{BERS01}. 
We denote the least
simply-local model of $P$ containing $\mathit{SM}_P$ by 
$\mathit{PM}^{SL}_P$, for \emph{partial model}.

 \begin{example}\label{ex:local-model}
  Consider \texttt{APPEND}. To compute
$\mathit{PM}^{SL}_{\tt APPEND}$, we must iterate 
the abovementioned variant of the $T_P$-operator starting from 
the fact clause  `\verb:append([],Ys,Ys).:' and any 
 simply moded atom. It turns out that
\begin{eqnarray*}
\mathit{PM}^{SL}_{\tt APPEND} 
& =\;\; \displaystyle \bigcup_{n=0}^{\infty}\;\; ( \!\!\!\!\!\! & 
\{ {\tt append}([T_1,\ldots,T_n],T,[T_1,\ldots,T_n|T]) \} \cup \\ 
& &
\{ 
{\tt append}([T_1,\ldots,T_n|S],T,[T_1,\ldots,T_n|\mathtt{X}])\mid 
\mathtt{X} \mbox{ is fresh}\}).
\end{eqnarray*}
We refer to \cite{BERS01} for the details of this calculation.
\end{example}

\decrease
We now define {\em simply-acceptability}, which is the notion of
decrease used for proving input termination.

We write $p \simeq q$ if $p$ and $q$ are mutually recursive
predicates \cite{Apt97}. Abusing notation, we also use $\simeq$ for
{\em atoms}, where $p(\vect{s},\vect{t}) \simeq q(\vect{u},\vect{v})$
stands for $p \simeq q$.

\begin{definition}
  Let $P$ be a program, $|.|$ a moded generalised\footnote{
In \cite{BERS01}, the word  ``generalised'' is dropped, but 
here we prefer to emphasise that non-ground atoms are 
included in the domain.}
 level mapping and
  $I$ a simply-local model of $P$ containing $\mathit{SM}_P$. 
A clause \clausab\ is 
  \emph{simply-acceptable by
$|.|$ and $I$} if for
  every substitution $\theta$ simply-local w.r.t.~it,
\[ 
 \mbox{\rm for all}\ i \in [1,n], \quad
(B_1, \ldots, B_{i-1})\theta\in I \ \mbox{and} \
 A \simeq  B_i \quad
\mbox{implies} \quad |A\theta| > |B_i\theta|.
\]
The program $P$ is  \emph{simply-acceptable by  $|.|$ and  $I$} if 
each clause of $P$ is simply-acceptable by  $|.|$ and $I$.
\end{definition}

Admittedly, the proof obligations may be difficult to verify,
especially in the cases where a small (precise) simply-local model is
required. However, as our examples show, often it is not necessary at
all to consider the model, as one can show the decrease for arbitrary
instantiations of the clause.

\comparison
Unlike all other characterisations in this article,
simply-acceptability is not based on ground instances of clauses, but
rather on instances obtained by applying simply-local substitutions,
which arise in input-consuming derivations of simply moded
programs. This is also why we use {\em generalised} level mappings and 
a special kind of models. 

Also note that in contrast to recurrence and other decreasing notions
to be defined later, simply-acceptability has no proof obligation on
queries (apart from the requirement that queries must be simply
moded).  Intuitively, such a proof obligation is made redundant by the
mode conditions (simply-acceptability and moded level mapping) and the
fact that derivations must be input-consuming.  We also refer to
subsection~\ref{characterisations-subsec}.

\maintheorem
We can now show that this concept
allows to characterise the class of input terminating programs.

\begin{theorem}[\cite{BERS01}]
\label{input-termination-thm}
Let $P$ and $Q$ be a simply moded program and query.

If $P$ is simply-acceptable by some $|.|$ and $I$,
then $P$ and $Q$ input terminate.

Conversely, if $P$ and every simply moded query input
  terminate, then $P$ is simply-acceptable 
by some $|.|$ and $\mathit{PM^{SL}_P}$.
\end{theorem}

Note that the formulation of the theorem differs slightly
from the original for reasons of consistency, but one can 
easily see that the formulations are equivalent.

The definition of input-consuming derivations is
independent from the textual order of atoms in a query, and so the
textual order is irrelevant for termination. This means of course that 
if we can prove input termination for a program and query, we have
also proven termination for a program obtained by permuting the body
atoms of each clause and the query in an arbitrary way. This will be
seen in the next example. It would have been possible to state this
explicitly in the above theorem, but that would have complicated the
definition of simply-local substitution and subsequent definitions.
Generally, the question of whether or not it is necessary to make the
permutations of body atoms explicit is discussed in \cite{Sma99t}.

\examples
\begin{example}\label{ex:even-is-simply-acceptable}
The program {\tt EVEN} in
figure~\ref{even-prog} is simply-acceptable with modes 
$\tt even(\inp),\ lte(\out,\inp)$ by using the 
level mapping in Example~\ref{ex:even-query-is-not-recurrent},
interpreted as moded {\em generalised} level mapping in the obvious
way, 
and using any simply-local model. 
Moreover, the query 
$\tt even(X),\ lte(X,s^{100}(0))$ is permutation simply moded.
Hence {\tt EVEN} and this query input terminate.
\end{example}

\begin{example} \label{ex:permute-is-simply-acceptable}
Figure~\ref{permute-prog} shows the program {\tt PERMUTE}.
 Note that
$\tt permute \not\simeq insert$. Assume the modes
$\tt permute(\inp,\out)$, $\tt insert(\inp,\inp,\out)$. The program is
readily checked to be simply-acceptable, using the moded generalised level mapping
 \[ 
|\mathtt{permute}(Xs,Ys)| =  
|\mathtt{insert}(Xs,Ys,Zs)| = \mathit{size}(Xs)
\]
and any simply-local model. Thus the program and any simply moded
query input terminate.
It can also easily be shown that the program is not recurrent. 
\end{example}

\begin{figure}\figrule
\begin{minipage}[t]{5cm}
\begin{janprogram}
\%\> permute(Xs,Ys) \la \ \\
\%\> \> {\rm {\tt Ys} is a permutation of the list {\tt Xs}.}\\[2mm]
\> permute([X|Xs],Ys) \la \\
\>       \> permute(Xs,Zs), \\
\>       \> insert(Zs,X,Ys). \\
\> permute([],[]). 
\end{janprogram}
\end{minipage}
\hfill
\begin{minipage}[t]{5cm}
\begin{janprogram}
\%\> insert(Xs,X,Zs) \la \ \\
\%\> \> {\rm {\tt Zs} is obtained by inserting {\tt X} into {\tt Xs}}.\\[2mm]
\>  insert(Xs,X,[X|Xs]). \\
\>  insert([U|Xs],X,[U|Zs]) \la \\
\> \> insert(Xs,X,Zs). 
\end{janprogram}
\end{minipage}
\caption{\tt PERMUTE\label{permute-prog}}
\figrule\end{figure}

\begin{example}\label{ex:quicksort-is-simply-acceptable}
Figure~\ref{fig:qs} shows program 15.3 from \cite{SS86}: \texttt{QUICKSORT}
using a form of difference lists (we permuted two body atoms for the
sake of clarity). This program is simply moded with the modes
$\tt quicksort(\inp,\out)$, $\tt quicksort\_dl(\inp,\out,\inp)$, $\tt
partition(\inp,\inp,\out,\out)$, $\texttt{=<}(\inp,\inp)$, 
$\texttt{>}(\inp,\inp)$.

Let $|.|$ be the list-length function (see subsection~\ref{norms-subsec}). 
We use the following moded generalised level mapping  (positions with 
$\_$ are irrelevant)
\begin{eqnarray*}
|\mathtt{quicksort\_dl}(Xs,\_,\_)| &=& |Xs|,\\
|\mathtt{partition}(Xs,\_,\_,\_)| &=& |Xs|.
\end{eqnarray*}
The level mapping of all other atoms 
can be set to 0. Concerning the model, the simplest solution is to use 
the model that expresses the dependency between the list lengths of
the arguments of $\mathtt{partition}$, i.e., $I$ should
contain all atoms of the form
$\mathtt{partition}(S_1,X,S_2,S_3)$ where
$|S_1| \geq |S_2|$ and
$|S_1| \geq |S_3|$ ($|.|$ being the list-length function). 
Note that this includes all simply-moded atoms using
$\mathtt{partitition}$, and that this model is a fortiori 
simply-local since (\ref{SL-model-eq}) in
Definition~\ref{sl-model-def} is true even for arbitrary
$\theta$.

The program is then simply-acceptable by $|.|$ and $I$ and hence
input terminates for every simply moded query. 
\end{example}

\begin{figure}\figrule
\begin{program}
\% \> quicksort(Xs, Ys) \la Ys {\rm is an ordered permutation of } Xs.
\\[2mm]
 \>         quicksort(Xs,Ys) \la quicksort\_dl(Xs,Ys,[]).\\[2mm]
 \>         quicksort\_dl([X|Xs],Ys,Zs)  \la \\
 \> \>         partition(Xs,X,Littles,Bigs),\\
 \> \>         quicksort\_dl(Bigs,Ys1,Zs). \\ 
 \> \>         quicksort\_dl(Littles,Ys,[X|Ys1]),\\ 
 \>         quicksort\_dl([],Xs,Xs).\\[2mm]
 \>         partition([X|Xs],Y,[X|Ls],Bs) \la  X =< Y, partition(Xs,Y,Ls,Bs).\\
 \>         partition([X|Xs],Y,Ls,[X|Bs]) \la X > Y,  partition(Xs,Y,Ls,Bs).\\
 \>         partition([],Y,[],[]).
\end{program}
\caption{\texttt{QUICKSORT}\label{fig:qs}}
\figrule\end{figure}

\section{Local Delay Termination}\label{local-sec}
\limitation The class of programs and queries that terminate for all
input-consuming derivations is considerable, but there are still many
interesting programs not contained in it.

\begin{example} \label{ex:ic-strictly-lds}
  Consider again the {\tt PERMUTE} program 
  (figure~\ref{permute-prog}), but this time assume the mode 
$\tt permute(\out,\inp), \tt insert(\out, \out, \inp)$. 
Consider also the query $\tt permute(X,[1])$.  It is easy to check
  that there is an infinite {\em input-consuming} derivation for this
  query obtained by selecting always the leftmost atom that can be
  selected.  In fact, \texttt{PERMUTE} in this mode 
cannot be simply-acceptable,
not even after reordering of atoms in clause bodies.
To see this, we first reorder the body atoms of the recursive clause
to obtain 
\begin{janprogram}
\> permute([X|Xs],Ys) \la \\
\>       \> insert(Zs,X,Ys), \\
\>       \> permute(Xs,Zs). 
\end{janprogram}
so that the program is simply moded and thus our method
showing input termination is applicable in principle. Now
$\mathit{PM}^{SL}_\mathtt{PERMUTE}$ contains every atom of the form
$\mathtt{insert}(\mathtt{Us},\mathtt{U},\mathit{Vs})$, i.e., every simply moded 
atom whose predicate is $\mathtt{insert}$. Therefore in particular
$\mathtt{insert}(\mathtt{Us},\mathtt{U},\mathtt{Vs})
\in \mathit{PM}^{SL}_\mathtt{PERMUTE}$ (note that $\mathtt{Vs}$ is a
variable). The substitution 
$\theta = \{\tt Ys/Vs, Zs/Us, X/U \}$ is simply-local w.r.t.~the 
clause. Therefore, for the clause to be simply-acceptable, 
there would have to be a moded generalised level mapping such that 
$\tt |permute([U|Xs],Vs)| > |permute(Xs,Us)|$. This is a contradiction 
since a {\em moded} generalised level mapping is necessarily defined
as a generalised norm of the second argument of $\mathtt{permute}$, 
and $\mathtt{Vs}$ and $\mathtt{Us}$ are equivalent modulo variance.

However, all derivations for this query are finite w.r.t.~the RD
selection rule, which for this example happens to be an instance of
the selection rules considered in this section.
\end{example}

\termination
Marchiori and Teusink \shortcite{MT99} have considered local selection
rules controlled by delay declarations. They define a {\em safe delay
declaration} so that an atom can be selected only when it is bounded
w.r.t.~a level mapping. In order to avoid even having to define delay
declarations, we take a shortcut, by defining the following.

\begin{definition}\label{delay-safe-def}
  A selection rule is {\em delay-safe (w.r.t.~$|.|$)} if it specifies
  that an atom $A$ can be selected only when $A$ is bounded
  w.r.t.~$|.|$. 
\end{definition}

Note that delay-safe selection rules imply that the depth of the
SLD-tree of the selected atom does not depend on further instantiation
as in the previous section.

\begin{definition} \label{c2:def:ldst}
  A program $P$ and query $Q$ {\em local delay terminate}
  (w.r.t.~$|.|$) if they universally terminate w.r.t.~the set of
  selection rules that are both local and delay-safe (w.r.t.~$|.|$).
\end{definition}

\contrived Unlike in the previous section, modes are not used
explicitly in the definition of delay-safe selection rules.
Therefore it is possible to invent an example of a program and a query
that input terminate but do not local delay terminate.  Such an
example is of course contrived, in that the level mapping is chosen 
in an inappropriate way. 

\begin{example}\label{ex:ic-yes-lds-no}
  The {\tt APPEND} program and the query 
{\tt append([],[],X),\ append(X,[],Y)}
input terminate for the mode 
{\tt append(\inp,\inp,\out)}. However, they do not local delay
  terminate w.r.t.~a level mapping $|.|$ such that $|A| = 0$ for every
  $A$ (e.g., consider the  RD selection rule).
\end{example}

However, in section~\ref{relations-sec} we will see
that under natural assumptions (in particular, the
level mapping must be moded) delay-safe selection rules are
also input-consuming. Then, input termination implies
local delay termination, and as is witnessed by
Example~\ref{ex:ic-strictly-lds}, this implication is strict.

\onlyhere{Information on Data Flow: Covers}
 Delay-safe selection rules ensure that selected atoms are bounded.
To ensure that the level mapping {\em decreases} during a derivation,
we exploit additional information provided by a model of the program.
Given an atom $B$ in a query, we are interested in other atoms that
share variables with $B$, so that instantiating these variables makes
$B$ bounded. A set of such atoms is called a {\em direct cover}. The only way 
of making $B$ bounded is by resolving away one of its direct covers. 
The formal definition is as follows. 

\begin{definition}\label{cover-def}
  Let $|.|$ be a level mapping, $A \la Q$ a clause containing a body
  atom $B$, and $\tilde{C}$ a subset\footnote{By abuse of terminology,
here we identify a query with the set of atoms it contains.} 
of $Q$ such that $B \not \in
  \tilde{C}$. We say that $\tilde{C}$ is a {\em direct cover for $B$
  (w.r.t.~$A \la Q$ and $|.|$)} if there exists a substitution
  $\theta$ such that $B\theta$ is bounded w.r.t.~$|.|$ and
  $\dom(\theta) \subseteq \vars(A,\tilde{C})$.

  A direct cover is {\em minimal} if no proper subset is a direct
 cover.  
\end{definition}

Note that the above concept is similar to well-modedness, assuming a
moded level mapping. In this case, for each atom, the atoms to the
left of it are a direct cover. This generalises in the obvious way to
{\em permutation} well moded queries.  

Considering an atom $B$, we have said that the only way of making $B$
bounded is by resolving away one of $B$'s direct covers. However, for an
atom in a direct cover, say atom $A$, to be selected, $A$ must be
bounded, and the only way of making $A$ bounded is by resolving away
one of $A$'s direct covers. Iterating this reasoning gives rise to
a kind of closure of the notion of direct cover.

\begin{definition}
Let $|.|$ be a level mapping and $A \la Q$ a clause. 
Consider the least set $\cal C$, subset of 
${\mathcal P}(Q \times {\mathcal P}(Q))$, such that
\begin{enumerate}
\item
$\langle B, \emptyset \rangle \in {\cal C}$ whenever $B$ has
$\emptyset$ as minimal direct cover for $B$ in $A \la Q$;
\item
$\langle B,  \tilde{C} \rangle \in  {\cal C}$ whenever 
$B\not\in \tilde{C}$, and 
$\tilde{C} = \{C_1,\dots,C_k\}\cup \tilde{D}_1 \cup \dots \tilde{D}_k$, 
where $\{C_1,\dots,C_k\}$ is a minimal direct
cover of $B$ in $A \la Q$, and for $i \in [1,k]$, 
$\langle C_i,\tilde{D}_i\rangle \in  {\cal C}$.
\end{enumerate}
The set
$Covers(A \la Q) \subseteq Q \times {\mathcal P}(Q)$ 
is defined as the set obtained by deleting from ${\cal
C}$ each element of the form $\langle B, \tilde{C} \rangle$ if there
exists another element of ${\cal C}$ of the form $\langle B,
\tilde{C}' \rangle$ such that $\tilde{C}' \subset \tilde{C}$.

We say that $\tilde{C}$ is a {\em cover for $B$ (w.r.t.~$A \la Q$ and
$|.|$)} if $\langle B, \tilde{C} \rangle$ is an element of
$Covers(A \la Q)$.
\end{definition}

\decrease
 The following concept is used to show that programs terminate
for local and delay-safe selection rules. We present a definition slightly
different from the original one \cite{MT99}, albeit equivalent.

\begin{definition}\label{delay-recurrent}
  Let $|.|$ be a level mapping and $I$ a Herbrand interpretation. 
  A program $P$ is \emph{delay-recurrent by $|.|$ and $I$} if 
  $I$ is a model of $P$, and for
  every clause $\cequals A \la \tn{B}{n}$ of $P$, for every $i \in [1,n]$,
  for every cover $\tilde{C}$ for $B_i$, for every substitution
$\theta$ such that $c\theta$ is ground, 
\[       \mbox{if} \ \  I \Mo \tilde{C}\theta \ \ 
         \mbox{then} \ \ |A\theta| > |B_i\theta|.
\]
\end{definition}

 We believe that this notion should have better been called {\em
delay-acceptable}, since the convention is to call decreasing notions
that involve models {\em (\dots)-acceptable}, and the ones that do not
involve models {\em (\dots)-recurrent}.

The most essential
differences between delay-recurrence and simply-acceptability are that
the former is based on models, whereas the latter is based on
simply-local models, and that the former requires decreasing for
all body atoms, whereas the latter only for mutually recursive calls.

Just as simply-acceptability, delay-recurrence imposes no proof
obligation on queries. Such a proof obligation is made redundant by
the fact that selected atoms must be bounded. Note that if no most
recently introduced atom in a query is bounded, we obtain termination
by deadlock.
We also refer to subsection~\ref{characterisations-subsec}. 

\maintheorem 
In order for delay-recurrence to ensure termination, it is crucial
that when an atom is selected, its cover is resolved away completely
(this allows to use the premise $I \models \tilde{C}\theta$ in
Definition~\ref{delay-recurrent}). To this end, local selection rules must
be adopted. We can now state the result of this section.

\begin{theorem}[\cite{MT99}]\label{local-thm}
  Let $P$ be a program.

If $P$ is delay-recurrent by a level mapping $|.|$ and a Herbrand
  interpretation $I$, then for every query $Q$, $P$ and $Q$ local
  delay terminate.
\end{theorem}

\subsection{Example}
\begin{example}\label{permute-is-delay-recurrent-ex}
  Consider again {\tt PERMUTE} (figure~\ref{permute-prog}),
with the level mapping and model
\begin{eqnarray*}
 |\mathtt{permute}(xs,ys)|   & = & |ys| + 1\\[-0.5ex]
 |\mathtt{insert}(xs,ys,zs)| & = &  |zs|\\[0.5ex]
I                            & = &
  \{ \mathtt{permute}(xs,ys) \mid |xs| = |ys| \} \U \\[-0.5ex]
                             &   &
  \{ \mathtt{insert}(xs,y,zs) \mid  |zs| = |xs| + 1 \}.
\end{eqnarray*}
The program is delay-recurrent by $|.|$ and $I$. We check the
recursive clause for $\tt permute$. Consider an arbitrary ground instance
\begin{equation} \label{permuteexcit}
\mathtt{permute}([x|xs],ys) \la  \mathtt{permute}(xs,zs),\ 
\mathtt{insert}(zs,x,ys).
\end{equation}
First, we observe that $I$ is a model of this instance. In fact, if
its body is true in $I$, then 
$|ys| = |zs| + 1$ and 
$|xs| = |zs|$. This implies $|ys| = |xs| + 1$, and hence 
$\mathtt{permute}([x|xs],ys)$ is true in $I$.

Let us now show the decrease from the head to the $\tt permute$ body
atom. There is only one cover $\tt insert(Zs,X,Ys)$, so we must
show that
\[ 
|ys| = |zs| + 1 \quad \mbox{implies} \quad |ys| + 1 > |zs| + 1,
\]
which is clearly true.  Now consider the second body atom. It has
an empty cover. This time, for every instance of the clause such that
the head is ground, we have that $|ys| + 1 > |ys|$. Hence we have
shown that the clause is delay-recurrent.

It is interesting to compare this to Example~\ref{ex:ic-strictly-lds}, 
where we were not able to show a decrease.
\end{example}

\incompleteness
Note that delay-recurrence is a sufficient but not necessary condition
for local delay termination. The limitation lies in the notion of
cover: to make an atom bounded, one has to resolve one of its covers;
but conversely, it is not true that resolving any cover will make the
atom bounded.

\begin{example} \label{ex:delay-recursive-incomplete}
Consider the following simple program
\begin{janprogram}
\> z $\la$  p(X), q(X), r(X).\\
\> p(0).\\
\> q(s(X)) $\la$ q(X).\\
\> r(X).
\end{janprogram}
The program and any query $Q$ local delay terminate w.r.t.~the level
mapping: 
\[
\begin{array}{c}
|\mathtt{z}| = |\mathtt{p}(t)| = |\mathtt{r}(t)| = 0\\
|\mathtt{q}(t)| = size(t)
\end{array}
\]
In fact, the only source of non-termination for a query might be an atom
 {\tt q($X$)}. However, for any such atom selected
by a delay-safe selection rule, $X$ is a ground term. Hence the
recursive clause in the program cannot generate an infinite
derivation.  On the other hand, it is not the case that the program is
delay-recurrent. Consider, in fact, the first clause. Since {\tt r(X)}
is a cover for {\tt q(X)}, we would have to show for some $|.|'$ that
for every $t$:
\[ 
|\tt z|' > |\mathtt{q}(t)|'. 
\]
This is impossible, since delay-recurrence on the third clause implies
$|\mathtt{q}(\mathtt{s}^k(0))|' \geq k$ for any natural $k$.
\end{example}

\section{Left-Termination}\label{left-sec}
\limitation
In analogy to previous sections, we should start this section with an
example illustrating that the assumption of local delay-safe selection 
rules is sometimes too weak to ensure termination, and thereby
motivate the ``stronger'' assumption of the LD selection rule.  
Such an example can easily be given.

\begin{example}\label{ex:lds-strictly-left}
Consider the program
\begin{janprogram}
\> p $\la$  q, p.
\end{janprogram}
with query $\tt p$, where $\tt |p| = |q| = 0$. 
It left-terminates but does not local delay terminate. 
\end{example}

However, the example is somewhat artificial, and in fact, we believe
that assuming the LD selection rule is only slightly stronger than
assuming an arbitrary local delay-safe selection rule, as far as
termination is concerned. Nevertheless, there are several reasons for
studying this selection rule in its own right.  First, the conditions
for termination are easier to formulate than for local delay
termination. Secondly, the vast majority of works consider this rule,
being the standard selection rule of Prolog. Finally, for the class of
programs and queries that terminate w.r.t. the LD selection rule we
are able to provide a sound and complete characterisation.

\termination

\begin{definition} \label{c2:def:leftt}
A program $P$ and query $Q$ {\em left-terminate\/} if they
universally terminate w.r.t.~the set consisting of only the LD
selection rule.
\end{definition}

\contrived 
Formally comparing this class to the two previous
ones is difficult. In particular, 
left-termination is not necessarily
stronger than input or local delay termination.

\begin{example} \label{ex:lds-yes-left-no}
We have shown in Example~\ref{permute-is-delay-recurrent-ex} that 
{\tt PERMUTE} and every query local delay terminate w.r.t.~the level
mapping given there.  
Moreover, no derivation deadlocks. However, {\tt PERMUTE} 
and the query $\tt permute(X,[1])$ do not left
terminate. Similarly to Example~\ref{ex:ic-strictly-lds}, this example is
contrived since the program is  intended for the RD selection rule.

One could easily construct a similar example comparing left
termination with input termination. 
\end{example}

Also, local delay termination may not imply left-termination
because of the deadlock problem. 

\incompleteness 
Left-termination was addressed by Apt \& Pedreschi \shortcite{AP93r},
who introduced the class of acceptable logic programs.  However, their
characterisation encountered a completeness problem similar to the one
highlighted for Theorem~\ref{c2:rec:thm}.

\begin{example} \label{transp-ex}
Figure~\ref{transp-prog} shows {\tt TRANSP}, a program that terminates on 
a strict subset of ground queries only.
 In the intended meaning of the program, 
$\mathtt{trans}(x,y,e)$ succeeds iff 
$x \leadsto_e y$, i.e.~if $\mathtt{arc}(x,y)$ is
in the transitive closure of a direct acyclic graph (DAG) $e$, which
is represented as a list of arcs. It is readily checked that if $e$ is
a graph that contains a cycle, infinite derivations may occur.
\begin{figure}\figrule
\begin{minipage}[t]{5cm}
\begin{janprogram}
\%\> trans($x$,$y$,$e$) \la \mbox{\rm $x \leadsto_e y$ for a DAG $e$}\\[2ex]
\> trans(X,Y,E) \la\\
\> \>  member(arc(X,Y),E).\\[2ex]
\> trans(X,Y,E) \la\\
\> \>  member(arc(X,Z),E), trans(Z,Y,E).
\end{janprogram}
\end{minipage}
\hfill
\begin{minipage}[t]{5cm}
\begin{janprogram}
\\[2ex]
\> member(X,[X|Xs]).\\[2ex]
\> member(X,[Y|Xs]) \la\\
\> \>  member(X,Xs).
\end{janprogram}
\end{minipage}
\caption{\tt TRANSP\label{transp-prog}}
\figrule\end{figure}

In the approach by Apt \& Pedreschi, {\tt TRANSP} cannot be reasoned
about, since the same incompleteness problem as for recurrent programs
occurs, namely that they characterise a class of programs that 
(left-)terminate for every ground query.  
\end{example}

 The cause of the restricted form of completeness of
Theorem~\ref{c2:rec:thm} lies in the use of level mappings, which must
specify a natural number for every ground atom --- hence termination is
forced for every ground query.  A more subtle problem with using level
mappings is that one must specify values also for {\em uninteresting
  atoms}, such as $\mathtt{trans}(x,y,e)$ when $e$ is not a DAG.
The solution to both problems is to consider 
{\em extended} level mappings~\cite{Rug97t,Rug99}.

\begin{definition} \label{c2:def:lm}
  An {\em extended level mapping\/} is a function $|.|:B_L \ra
  \nat^{\infty}$ of ground atoms to $\nat^{\infty}$, where
  $\nat^{\infty} = \nat \U \{ \infty \}$. 
\end{definition}

 The inclusion of $\infty$ in the codomain is intended to model
non-termination and uninteresting instances of program clauses.
First, we extend the $>$ order on $\nat$ to a relation $\ma$ on
$\nat^{\infty}$.

\begin{definition} \index{$\ma$}
  We define $n \ma m$ for $n, m \in \nat^{\infty}$ iff $n = \infty$ or
  $n > m$.  We write $n \maeq m$ iff $n \ma m$ or $n = m$.  
\end{definition}

\decrease
  Therefore, $\infty \ma m$ for every $m \in \nat^{\infty}$.
With this additional notation we are now ready to introduce (a revised
definition of) acceptable programs and queries.  A program $P$ is
acceptable \index{acceptable programs} if for every ground instance of
a clause from $P$, the level of the head is greater than the level of
each atom in the body such that the body atoms to its left are true in
a Herbrand model of the program.

\begin{definition} \label{c2:def:accp}\label{c2:def:accq} 
  Let $|.|$ be an extended level mapping, and $I$ a Herbrand
  interpretation.  A program $P$ is \emph{acceptable by $|.|$ and
    $I$} if $I$ is a model of $P$, and for every \clausab\ in
  $\groun{L}{P}$:
\[ 
 \mbox{\rm for all}\ i \in [1,n], \quad
           I \Mo B_1, \ldots, B_{i-1} \quad 
\mbox{implies} \quad |A|  \ma |B_i|.
\]
A query $Q$ is {\em acceptable by $|.|$ and $I$\/} if there exists 
$k \in \nat$ such that for every $\tn{A}{n}$ $\in$  $\groun{L}{Q}$:
\[ 
\mbox{\rm for all}\ i \in [1,n], \quad  
    I \Mo A_1, \ldots, A_{i-1} \quad 
\mbox{implies} \quad k\ \ma |A_i|.  
\]
\end{definition}

\comparison 
Let us compare this definition with the
definition of delay-recurrence (Definition~\ref{delay-recurrent}). In
the case of local and delay-safe selection rules, an atom cannot be
selected before one of its covers is completely resolved. In the case
of the LD selection rule, an atom cannot be selected before the atoms
to its left are completely resolved. Because of the correctness of LD
resolution~\cite{Apt97}, this explains why, in both cases, a decrease
is only required if the instance of the cover, resp.~the instance of
the atoms to the left, are in some model of the program.  We also
refer to subsection~\ref{characterisations-subsec}.

\maintheorem 
Acceptable programs and queries precisely characterise left-termination. 

\begin{theorem}[\cite{AP93r,Rug97t}] \label{c2:acc:tcor} 
Let $P$ be a program and $Q$ a query.

If $P$ and $Q$ are both acceptable by an extended level mapping $|.|$
  and a Herbrand interpretation $I$, then $P$ and $Q$ left-terminate.

Conversely, if $P$ and $Q$ left-terminate, then there exist
an extended level mapping $|.|$ and a Herbrand interpretation $I$
such that $P$ and $Q$ are both acceptable by $|.|$ and $I$.  
\end{theorem}

\subsection{Example}
\begin{example} \label{ex:transp-is-acceptable}
We will show that {\tt TRANSP} is acceptable.  We have pointed out
that in the intended use of the program, $e$ is supposed to be a
DAG. We define:
\begin{eqnarray*}
|\mathtt{trans}(x,y,e)| & = & 
                    \left\{ 
                       \begin{array}{ll}
                        |e| + 1 + Card \{ v \mid x \leadsto_e v  \} 
                           &  \mbox{if $e$ is a DAG}\\
                        \infty & \mbox{otherwise}\\
                       \end{array} 
                    \right. \\[-0.5ex]
|\mathtt{member}(x,e)| & = & |e|\\[0.5ex]
I                             & = &  
\{ \mathtt{trans}(x,y,e) \mid x, y, e \in U_L \} \U \\[-0.5ex]
                              &   &        
\{ \mathtt{member}(x,e) \mid  x\ \mbox{is in the list}\ e \}.
\end{eqnarray*}
where $Card$ is the set cardinality operator. It is easy to
check that {\tt TRANSP} is acceptable by $|.|$ and $I$. In
particular, consider a ground instance of the second clause:
\[
\mathtt{trans}(x,y,e) \la
  \mathtt{member}(\mathtt{arc}(x,z),e), \mathtt{trans}(z,y,e).
\]
It is immediate to see that $I$ is a model of it.  In addition, we
have the proof obligations:
\begin{eqnarray*}
\mbox{\it (i)\/}& & |\mathtt{trans}(x,y,e)| \ma 
        |\mathtt{member}(\mathtt{arc}(x,z),e)|\\
\mbox{\it (ii)\/} & \mathtt{arc}(x,z)\ \mbox{is in}\ e \ \Ra &
|\mathtt{trans}(x,y,e)| \ma 
        |\mathtt{trans}(z,y,e)|.
\end{eqnarray*}
The first one is easy to show since 
$|\mathtt{trans}(x,y,e)| \ma |e|$. 
Considering the second one, we distinguish two cases.
If $e$ is not a DAG, the conclusion is immediate. Otherwise, 
$\mathtt{arc}(x,z)$ in $e$ implies that
$Card \{ v \mid x \leadsto_e v \} >                            
 Card \{ v \mid z \leadsto_e v \}$, and so:
\begin{eqnarray*}
|\mathtt{trans}(x,y,e)| 
& =   & |e| + 1 + Card \{ v \mid x \leadsto_e v \} \\
& \ma & |e| + 1 + Card \{ v \mid z \leadsto_e v \} = 
|\mathtt{trans}(z,y,e)|.
\end{eqnarray*}
Finally, observe that for a DAG $e$, the queries 
$\mathtt{trans}(x,\mathtt{Y},e)$ and 
$\mathtt{trans}(\mathtt{X,Y},e)$ are acceptable by $|.|$ and $I$. 
The first one is intended to compute all nodes $y$ such
that $x \leadsto_e y$, while the second one computes the binary
relation $\leadsto_e$.
Therefore, the {\tt TRANSP} program and those queries 
left-terminate.

\label{ex:strong-strictly-left}
Note that this is of course also an example of a program and a query which
left-terminate but do not strongly terminate (e.g., consider the RD
selection rule).
\end{example}

\section{$\exists$-Termination} \label{c2:sec:ex}\label{exists-sec}
\limitation
So far we have considered four classes of terminating programs, making
increasingly strong assumptions about the selection rule, or in other
words, considering in each section a smaller set of selection rules. In the
previous section we have arrived at a singleton set containing the LD
selection rule. Therefore we can clearly not strengthen our
assumptions, in the same sense as before, any further. 

We will now consider an assumption about the selection rule which is
equally abstract as assuming {\em all} selection rules
(section~\ref{strong-sec}).
We introduce {\em $\exists$-termination\/} of logic
programs, claiming that it is an essential concept for separating the
{\em logic\/} and {\em control} aspects of a program.

Before, however, we motivate the limitations of left-termination.

\begin{example} \label{ex:left-strictly-exists} 
The program {\tt PRODCONS} in figure~\ref{prodcons-prog} 
abstracts a (concurrent) system
composed of a producer and a consumer.
For notational convenience, we identify the term {\tt s$^n$(0)} with
the natural number $n$.  Intuitively, {\tt prod} is the producer of a
non-deterministic sequence of $1$'s and $2$'s, and {\tt cons} the
consumer of the sequence.  The shared variable {\tt Bs} in clause 
{\it (s)\/} acts as an unbounded buffer.  
The overall system is started by the query $\mathtt{system}(n)$.  Note
that the program is well moded with the obvious mode 
$\{\tt prod(\out), cons(\inp,\inp), wait(\inp)\}$, but assuming LD
(and hence, input-consuming) derivations does not ensure
termination. The crux is that {\tt prod} can produce a message
sequence of arbitrary length. Now {\tt cons} can only consume a message
sequence of length $n$, but for this to ensure termination, atoms
using $\tt cons$ must be eventually selected. 
We will see that a selection rule exists for which this program and
the query $\mathtt{system}(n)$ terminate.
\end{example}

\begin{figure}\figrule
\begin{minipage}[t]{5cm}
\begin{programma}
{\it (s)\/}\> system(N) \la\\
\> \> prod(Bs),  cons(Bs,N).\\[2mm]
{\it (p1)\/}\> prod([s(0)|Bs])) \la \\
\> \> prod(Bs).\\
{\it (p2)\/}\> prod([s(s(0))|Bs])) \la \\
\> \> prod(Bs).\\
{\it \/}\> prod([]).
\end{programma}
\end{minipage}
\hfill
\begin{minipage}[t]{5cm}
\begin{programma}
{\it (c)\/}\> cons([D|Bs],s(N))  \la \\
\> \> cons(Bs,N), wait(D).\\
{\it \/}\> cons([], 0).\\[2mm]
{\it (w)\/}\> wait(s(D)) \la \\
\> \> wait(D). \\
{\it \/} \> wait(0).
\end{programma}
\end{minipage}
\caption{\tt PRODCONS\label{prodcons-prog}}
\figrule\end{figure}

\termination
We introduce next the notion of $\exists$-termination.

\begin{definition}\label{c2:def:eut}
  A program $P$ and a query $Q$ $\exists$-terminate if there exists a
  non-empty set ${\cal S}$ of standard selection rules such that $P$
  and $Q$ universally terminate w.r.t.~$\cal S$.  
\end{definition}

 If $P$ and $Q$ do not $\exists$-terminate, then
no standard selection rule can be terminating.  For extensions of the
standard definition of selection rule, such as input-consuming and
delay-safe rules, this is not always true.

\begin{example} \label{ex:ic-or-lds-not-imply-exists}
  The simple program
\begin{janprogram}
\> p(s(X)) $\la$  p(X).\\
\> p(X).
\end{janprogram}
with mode $\tt p(\inp)$ and query $\tt p(X)$ input
terminates, but does not $\exists$-terminate. 
The same program and query 
local delay terminate (w.r.t.~$|${\tt p($t$)}$| = size(t)$).  
\end{example}

 In section~\ref{relations-sec}, we will show that {\em permutation
  well-modedness} is a sufficient condition to ensure that if $P$ and
$Q$ input terminate then they $\exists$-terminate.

Here, we observe that $\exists$-termination coincides with
universal termination w.r.t. the set of fair selection rules.
Therefore, any fair selection rule is a terminating control for any
program and query for which a terminating control exists.

\begin{theorem}[\cite{RugTCS98,Rug99}] \label{c2:fair:unifair}
A program $P$ and a query $Q$ $\exists$-terminate
iff they universally terminate w.r.t.~the set of fair selection rules.
\end{theorem}

Concerning Example~\ref{ex:left-strictly-exists}, it can be said that 
viewed as a concurrent system, the program inherently
relies on fairness for termination.

\decrease
 Ruggieri \shortcite{RugTCS98,Rug99} offers a characterisation of
$\exists$-termination using the notion of {\em fair-bounded\/}
programs and queries. Just as
Definition~\ref{c2:def:accp}, it is based on {\em extended} level mappings. 

\begin{definition} \label{c2:def:fairp} \label{c2:def:fairq}
 Let $|.|$ be an extended level mapping, and $I$ a Herbrand interpretation.  A
program $P$ is \emph{fair-bounded by $|.|$ and $I$} if $I$ is
a model of $P$ such that for every $\clausab$ in $\groun{L}{P}$:
\begin{enumerate} 
\item[{\em (a)\/}] 
$I \Mo \tn{B}{n} 
\ \mbox{\rm implies that for every}\ 
i \in [1,n],\ 
|A| \ma |B_i|$, and
\item[{\em (b)\/}]  
$I \not \models \tn{B}{n} 
\ \mbox{implies that there exists}\
i \in [1,n] 
\ \mbox{with}\ 
I \not \models B_i \A |A| \ma |B_i|$.
\end{enumerate}

A query $Q$ is {\em fair-bounded by $|.|$ and $I$\/} if there exists
$k \in \nat$ such that for every $\tn{A}{n} \in \groun{L}{Q}$:
\begin{enumerate}
\item[{\em (a)\/}] 
$I \Mo A_1, \ldots, A_n  
\ \mbox{\rm implies that for every}\ 
i \in [1,n],\ 
k  \ma |A_i|$, and
\item[{\em (b)\/}]  
$I \not \models A_1, \ldots, A_n 
\ \mbox{implies that there exists}\
i \in [1,n] 
\ \mbox{with}\ 
I \not \models A_i \A k \ma |A_i|$. 
\end{enumerate}
\end{definition}

 Note that the hypotheses of conditions {\it (a)\/} and {\it (b)\/}
are {\em mutually exclusive\/}.

\comparison 
Let us discuss in more detail the meaning of proof obligations {\it
(a)\/} and {\it (b)\/} in Definition~\ref{c2:def:fairp}. Consider a
ground instance $\clausab$ of a clause.

If the body \tn{B}{n} is true in the model $I$, then there might exist a
SLD-refutation for it. Condition {\it (a)\/}  is then intended to bound the
length of the refutation.

If the body is not true in the model $I$, then it cannot have a
refutation. In this case, termination actually means that there is an
atom in the body that has a finitely failed SLD-tree.  
Condition {\it (b)\/} is then intended to bound the depth of the finitely failed
SLD-tree.  As a consequence of this, the complement of $I$ is
necessarily included in the finite failure set of the program.

Compared to acceptability, the model and the extended level mapping in
the proof of fair-boundedness have to be chosen more carefully, due to
more binding proof obligations.  As we will see in
section~\ref{relations-sec}, however, the simpler proof obligations of
recurrence and acceptability are sufficient conditions for proving
fair-boundedness. 
Note also that, as in the case of
acceptable programs, the inclusion of $\infty$ in the codomain of
extended level mapping allows for excluding {\em unintended atoms\/}
and {\em non-terminating atoms\/} from the termination analysis.  In
fact, if $|A| = \infty$ then {\it (a, b)\/} in Definition~\ref{c2:def:fairp}
are trivially satisfied.

\maintheorem
 Fair-bounded programs and queries precisely characterise
 $\exists$-termination, i.e.~the class of logic programs and queries
 for which a terminating control exists.

\begin{theorem}[\cite{RugTCS98,Rug99}] \label{c2:fair:ncscorr}
Let $P$ be a program and $Q$ a query.

If $P$ and $Q$ are both fair-bounded by an extended level mapping $|.|$
  and a Herbrand interpretation $I$, then $P$ and $Q$ $\exists$-terminate.

Conversely, if $P$ and $Q$ $\exists$-terminate, then there exist
an extended level mapping $|.|$ and a Herbrand interpretation $I$
such that $P$ and $Q$ are both fair-bounded by $|.|$ and $I$.  
\end{theorem}

\subsection{Example}

\begin{example} \label{ex:prodcons-is-fair-bounded} 
The {\tt PRODCONS} program is fair-bounded. First, we introduce the
{\em list-max\/} norm: \index{$lmax()$}
\[
\begin{array}{rcll}
lmax( f ( x_1 , \LL , x_n ) ) & = & 0 
& \mbox{if} \ f \neq [ \: . \: | \: . \: ] \\
lmax([x | xs]) & = & 
max\{ lmax( xs ), size(x) \} 
& \mbox{otherwise.}
\end{array}
\]
Note that for a ground list $xs$, $lmax(xs)$ equals the maximum size
of an element in $xs$.  Then we define:
\begin{eqnarray*}
| \mathtt{system}(n) | & = & size(n) + 3  \\[-0.5ex] 
| \mathtt{prod}(bs) |  & = & |bs| \\ [-0.5ex]
| \mathtt{cons}(bs,n) | & = & 
    \left\{ \begin{array}{ll} 
      size(n) + lmax(bs) & \mbox{ if } I\models \mathtt{cons}(bs,n) \\
      size(n)            & \mbox{ if } I\not \models \mathtt{cons}(bs,n) 
    \end{array} \right.\\[-0.5ex]
| \mathtt{wait}(t) | & = & size(t)\\[0.5ex]
I & = &    \{ \mathtt{system}(n) \mid n \in U_L  \}
        \U \{ \mathtt{prod}(bs) \mid lmax(bs) \leq 2  \} \U \\[-0.5ex]
  &   &    \{ \mbox{\tt cons($bs$,$n$)} \mid|bs| = size(n)  \}  
        \U \{ \mathtt{wait}(x) \mid x \in U_L  \}.
\end{eqnarray*}
Let us show the proof obligations of Definition~\ref{c2:def:fairp}. Those for
unit clauses are trivial. Consider now the recursive clauses 
{\it (w), (c), (p1), (p2),\/} and {\it (s)\/}.

\paragraph{\it (w).}
$I$ is obviously a model of {\it (w)\/}. In
  addition, 
$|\mathtt{wait(s}(d))|$ $=$ 
$size(d) + 1$ $\ma$ 
$size(d)$ $=$ 
$|\mathtt{wait}(d)|$. 
This implies {\it (a, b)\/}.

\paragraph{\it (c).}
Consider a ground instance 
$\mathtt{cons}([d|bs],\mathtt{s}(n)) \la 
\mathtt{cons}(bs,n),\ \mathtt{wait}(d)$
of {\it (c)\/}. If 
$I \Mo \mathtt{cons}(bs,n),\ \mathtt{wait}(d)$, then
$|bs| = size(n)$, and so
\[
|[d|bs]| = |bs| + 1 = size(n) + 1
= size(\mathtt{s}(n)),
\] 
i.e.~$I \Mo \mathtt{cons}([d|bs],\mathtt{s}(n))$.
Therefore, $I$ is a model of {\it (c)\/}. Let us show proof
obligations {\it (a, b)\/} of Definition~\ref{c2:def:fairp}.

\begin{itemize}
\item[{\it (a)\/}] 
Suppose that 
$I \models \mathtt{cons}(bs,n),\ \mathtt{wait}(d)$. 
We have already shown that 
$I$ $\models$ $\mathtt{cons}([d|bs],\mathtt{s}(n))$. 
We calculate:
\begin{eqnarray*}
|\mathtt{cons}([d|bs],\mathtt{s}(n))| 
& = & size(n) + 1 + max\{ lmax(bs), size(d) \} \\
& \ma & size(n) + lmax(bs) =  |\mathtt{cons}(bs,n)|\\
      |\mathtt{cons}([d|bs],\mathtt{s}(n))|   
& = & size(n) + 1 + max\{ lmax(bs), size(d) \} \\
& \ma & size(d)   
  =   |\mathtt{wait}(d)|.
\end{eqnarray*}
These two inequalities show that {\it (a)\/} holds. 

\item[{\it (b)\/}] 
If 
$I \not \models \mathtt{cons}(bs,n),\ \mathtt{wait}(d)$, 
then necessarily
$I \not \models \mathtt{cons}(bs,n)$. Therefore 
\begin{eqnarray*}
      |\mathtt{cons}([d|bs],\mathtt{s}(n))|   
& \maeq & size(n) + 1 \\
& \ma &   size (n) = 
      |\mathtt{cons}(bs,n)|,
\end{eqnarray*}
and so we have {\it (b)\/}. 
Recall that {\it (b)\/} states that the
depth of the finitely failed SLD-tree must be bounded. In fact, it is
the decrease of the ``counter'', the second argument of 
$\mathtt{cons}$, which in this case bounds the depth of the SLD-tree.
\end{itemize}

\paragraph{\it (p1,p2).}
$I$ is obviously a model of {\it (p1)\/}. Moreover we have
\[ 
|\mathtt{prod}([\mathtt{s}(0)|bs])| =  |bs| + 1 
\ \ma \ |bs| 
\ = \ |\mathtt{prod}(bs)|, 
\]
which implies {\it (a)\/} and {\it (b)\/}. The reasoning for {\it (p2)\/} 
is analogous.

\paragraph{\it (s).}
Consider a ground instance 
$\mathtt{system}(n) \la \mathtt{prod}(bs),\ \mathtt{cons}(bs,n)$
of {\it (s)\/}. 
Obviously $I$ is a model of {\it (s)\/}. Let us show
{\it (a,b)\/}.

\begin{itemize}
\item[{\it (a)\/}] Suppose that 
$I \Mo \mathtt{prod}(bs),\ \mathtt{cons}(bs,n)$.  
This implies $lmax(bs) \leq 2$ and $|bs|$ $=$
    $size(n)$.  These imply:
\begin{eqnarray*} \label{c2:lb:fin2}
|\mathtt{system}(n)| =  size(n) + 3 
& \ma & 
|bs| =   |\mathtt{prod}(bs)|\\
|\mathtt{system}(n)| =  size(n) + 3 
& \ma & 
size(n) + lmax(bs) =  |\mathtt{cons}(bs,n)|.
\end{eqnarray*}
These two inequalities show {\it (a)\/}.
\item[{\it (b)\/}] Suppose that $I \not \models \mathtt{prod}(bs),\ 
  \mathtt{cons}(bs,n)$.  Intuitively, this means that
  $\mathtt{prod}(bs)$, 
  $\mathtt{cons}(bs,n)$ has no refutation. We distinguish two cases.
  If $I \not \models \mathtt{cons}(bs,n)$ ($\mathtt{cons}(bs,n)$
  has no refutation) then:
\[
|\mathtt{system}(n)| \ = \ size(n) + 3 \ \ma \ size(n) \ = \ 
|\mathtt{cons}(bs,n)|,\] i.e. the depth of the SLD-tree of
$\mathtt{cons}(bs,n)$ is bounded (hence, the SLD-tree is finitely
failed). If $I \models \mathtt{cons}(bs,n)$ and $I \not \models
\mathtt{prod}(bs)$ ($\mathtt{prod}(bs)$
  has no refutation) then $|bs| = size(n)$, which implies:
\begin{eqnarray*} \label{c2:lb:fin4}
|\mathtt{system}(n)| 
 =  size(n) + 3 \ma |bs| = |\mathtt{prod}(bs)|,
\end{eqnarray*}
i.e. the depth of the SLD-tree of $\mathtt{prod}(bs)$ is bounded.
\end{itemize}

We conclude this example by noting that for every $n \in \nat$ the query
{\tt system($n$)} is fair-bounded by $|.|$ and $I$, and so
every fair SLD-derivation of {\tt PRODCONS} and 
$\mathtt{system}(n)$ is finite.
\end{example}

\section{Bounded Nondeterminism} \label{c2:sec:bn}\label{nondet-sec}
\limitation
In the previous section, we have made the strongest possible
assumption about the selection rule, in that we considered programs
and queries for which there {\em exists} a terminating control. In
general, a terminating control may not exist. Even in this case
however, all is not lost. If we can establish that a program and query
have only finitely many successful derivations, then we can transform
the program so that it terminates.

\begin{example} \label{ex:exists-strictly-bounded} 
The program {\tt ODDEVEN} in figure~\ref{oddeven-prog} 
defines the {\tt even} and {\tt odd} predicates,
with the usual intuitive meaning. The query 
$\, \mathtt{even(X),\; odd(X)}\, $ is
intended to check whether there is a number that is both even and odd.
It is readily checked that {\tt ODDEVEN} and the query do not
$\exists$-terminate. However, 
{\tt ODDEVEN} and the query have only finitely many, namely 0,
successful derivations. 
\end{example}
 
\begin{figure}\figrule
\begin{minipage}[t]{5cm}
\begin{janprogram}
\%\>  even(X) \la \ \\
\%\> \> {\tt X} {\rm is an even natural number.} \\[2mm]
\> even(s(X)) $\la$ odd(X).\\
\> even(0).
\end{janprogram}
\end{minipage}
\hfill
\begin{minipage}[t]{5cm}
\begin{janprogram}
\%\>  odd(X) \la \ \\
\%\> \> {\tt X} {\rm is an odd natural number.} \\[2mm]
\> odd(s(X)) $\la$ even(X).
\end{janprogram}
\end{minipage}
\caption{\tt ODDEVEN \label{oddeven-prog}}
\figrule\end{figure}

\termination
  Pedreschi \& Ruggieri \shortcite{PR99} propose the notion of
  {\em bounded nondeterminism\/} to model
  programs and queries with finitely many refutations.

\begin{definition} \label{c2:bn:bndef}
  A program $P$ and query $Q$ have {\em bounded nondeterminism} if
  for every standard selection rule $s$ there are finitely many
  SLD-refutations of $P$ and $Q$ via $s$.  
\end{definition}

 By the Switching Lemma \cite{Apt97}, each refutation via some
standard selection rule is isomorphic to some refutation via any other
standard selection rule. Therefore, bounded nondeterminism could have
been defined by requiring finitely many SLD-refutations of $P$ and $Q$
via {\em some} standard selection rule. Also, note that, while bounded
nondeterminism implies that there are finitely many refutations also
for non-standard selection rules, the converse implication does not
hold, in general (see Example~\ref{ex:ic-or-lds-not-imply-exists}).

Bounded nondeterminism, although not being a notion of termination in
the strict sense, is closely related to termination. In
fact, if $P$ and $Q$ $\exists$-terminate, then $P$ and $Q$ have
bounded nondeterminism. Conversely, if $P$ and $Q$ have bounded
nondeterminism then there exists an upper bound for the length of the
SLD-refutations of $P$ and $Q$. If the upper bound is known, then we
can syntactically transform $P$ and $Q$ into an equivalent program and
query that strongly terminate, i.e.~any selection rule will be a
terminating control for them.  Note that this transformation is even
interesting for programs and queries that $\exists$-terminate, since
few existing systems adopt fair selection rules. In addition, even if
we adopt a selection rule that ensures termination, we may apply the 
transformation to prune the SLD-tree from unsuccessful branches.

\decrease
In the following, we present a declarative characterisation of
programs and queries that have bounded nondeterminism, by introducing
the class of {\em bounded\/} programs and queries.
Just as Definitions~\ref{c2:def:accp} and~\ref{c2:def:fairp}, 
it is based on {\em extended} level mappings. 

\begin{definition}\label{c2:def:bfcp} \label{c2:def:bq} \index{bounded programs}
Let $|.|$ be an extended level mapping, and $I$ a Herbrand interpretation.  A
program $P$ is \emph{bounded by $|.|$ and $I$} if $I$ is a
model of $P$ such that for every $\clausab$ in $\groun{L}{P}$:
\[ 
I \Mo \tn{B}{n} 
\ \mbox{\rm implies that for every}\ 
i \in [1,n],\ 
|A| \ma |B_i|.
\]
A query $Q$ is {\em bounded by $|.|$ and $I$\/} if
there exists $k \in \nat$ such that for every $\tn{A}{n} \in
\groun{L}{Q}$:
\[ 
I \Mo A_1, \ldots, A_n  
\ \mbox{\rm implies that for every}\ 
i \in [1,n],\ 
k  \ma |A_i|.
\]
\end{definition}

\comparison  
It is straightforward to check that the definition of bounded programs
is a simplification of Definition~\ref{c2:def:fairp} of fair-bounded
programs, where proof obligation {\it (b)\/} is discarded.
Intuitively, the definition of boundedness only requires the
decreasing of the extended level mapping when the body atoms are true
in some model of the program, i.e.~they might have a refutation.

\maintheorem
Bounded programs and queries precisely characterise the
notion of bounded nondeterminism.

\begin{theorem}[\cite{PR99,Rug99}]
Let $P$ be a program and $Q$ a query.

If $P$ and $Q$ are both bounded by an extended level mapping $|.|$ and
a Herbrand interpretation $I$, then $P$ and $Q$ have bounded
nondeterminism.
  
Conversely, if $P$ and $Q$ have bounded nondeterminism, then there exist
an extended level mapping $|.|$ and a Herbrand interpretation $I$
such that $P$ and $Q$ are both bounded by $|.|$ and $I$.  
\end{theorem}

\examples
\begin{example} \label{oddeven-is-bounded}
Consider again the {\tt ODDEVEN} program. It is readily checked that
it is bounded by defining:
\begin{eqnarray*}
|\mathtt{even}(x)| =  |\mathtt{odd}(x)| & = & size(x)\\[-0.5ex]
I & = & 
\{ 
\mathtt{even}(\mathtt{s}^{2\cdot i}(0)),\  
\mathtt{odd}(\mathtt{s}^{2\cdot i + 1}(0)) \mid i \geq 0  \}.
\end{eqnarray*}
The query
$\,\mathtt{even(X),\; odd(X)}\,$ is bounded by $|.|$ and $I$. In fact, since no
instance of it is true in $I$, Definition~\ref{c2:def:bq} imposes no
requirement. Therefore, {\tt ODDEVEN} and the query above have bounded
nondeterminism.
\end{example}

Generally, for a query that has no instance in a model of the program
(it is {\em unsolvable}), the $k$ in Definition~\ref{c2:def:bfcp} can
be chosen as $0$. An automatic method to check whether a query (at a
node of a SLD-tree) is unsolvable has been proposed by \cite{BVWD98}.
Of course, the example is somewhat a limit case, since one does not
even need to run a query if it has been shown to be unsolvable. However,
we have already mentioned that the benefits of characterising bounded
nondeterminism also apply to programs and queries belonging to the
previously introduced classes.  In addition, it is still possible to
devise an example program and a {\em satisfiable} query that do not
$\exists$-terminate but  have bounded nondeterminism.

\begin{example} \label{all-is-bounded}
 We now define the predicate
  {\tt all} such that the query 
$\mathtt{all}(n_0,n_1,\mathtt{Xs})$ collects in
  {\tt Xs} the answers of a query
$\mathtt{q}(m,A)$ for
  values $m$ ranging from $n_0$ to $n_1$.
\begin{janprogram}
\> all(N,N,[A]) $\la$ q(N,A).\\
\> all(N,N1,[A|As]) $\la$ q(N,A), all(s(N),N1,As).\\
\> q(Y, Y). \%just as an example
\end{janprogram}
\end{example}

The program and the query {\tt all(0,s(s(0)),As)} do not
$\exists$-terminate, but they have only one computed answer, namely
{\tt As = [0,s(0),s(s(0))]}. The program and the query are bounded
(and thus have bounded nondeterminism) by defining:
\begin{eqnarray*}
|\mathtt{all}(n, m, x)| & = &  max\{ size(m) - size(n), 0 \} + 1\\[-0.5ex]
|\mathtt{q}(x,y)|       & = &  0 \\[0.5ex]
I & = & 
\{ 
\mathtt{all}(n, m, x) \mid size(n) \leq size(m) \} \U \\[-0.5ex]
  & = & \{\mathtt{q}(x,y)\}.
\end{eqnarray*}

\section{Relations between Classes} \label{c2:sec:class}\label{relations-sec}
We have introduced six classes of programs and queries, which
provide declarative characterisations of operational notions of
universal termination and bounded nondeterminism. In this section we 
summarise the relationships between these classes.

\subsection{Comparison of Characterisations}\label{characterisations-subsec}
We now try to provide an intuitive understanding of the significance
of the technical differences between the characterisations of
termination we have proposed. These are summarised in Table~\ref{char-table}.

\begin{table}
\normalsize
\caption{Comparison of characterisations\label{char-table}}
\begin{tabular}{rc@{\hspace{1.7em}}c@{\hspace{1.7em}}c@{\hspace{1.7em}}c@{\hspace{1.7em}}c@{\hspace{1.7em}}cc}
\hline\\[4.5ex]
                        & \begin{rotate}{25} only ground? \end{rotate} & 
                          \begin{rotate}{25} only recursive?\end{rotate} & 
                          \begin{rotate}{25} uses model? \end{rotate} & 
                          \begin{rotate}{25} query oblig.? \end{rotate} & 
                          \begin{rotate}{25} $\infty$ in codomain?\end{rotate} & 
                          \begin{rotate}{25} neg.~model info.?\end{rotate} & 
                          \hspace*{4em}\\  \hline
boundedness             & yes           & no                    & yes           & yes           & yes   & no\\ 
fair-boundedness        & yes           & no                    & yes           & yes           & yes   & yes\\ 
acceptability           & yes           & no                    & yes           & yes           & yes   & no\\ 
delay-recurrence        & yes           & no                    & yes           & no            & no    & no\\ 
simply-acceptability    & no            & yes                   & yes           & no            & no    & no\\ 
recurrence              & yes           & no                    & no            & yes           & no    & n.a.\\ 
\hline
\end{tabular}
\end{table}

The first difference concerns the question of whether a decrease is
defined for all ground instances of a clause, or rather for instances
specified in some other way. All characterisations, except
simply-acceptability, require a decrease for all ground instances of a
clause. One cannot clearly say that this difference lies in the nature
of the termination classes themselves: the first characterisation of
input-termination by Smaus~\shortcite{Sma99} also required a decrease
for the ground instances of a clause, just as there are
characterisations of left-termination \cite{BCF94,SVB92} based on
generalised level mappings and hence non-ground instances of clauses.
However, one can say that our characterisation of input-termination
inherently relies on measuring the level of non-ground atoms, which
may change via further instantiation. Nevertheless, this instantiation
is not arbitrary: it is controlled by the fact that derivations are
input-consuming and the programs are simply moded. This is reflected
in the condition that a decrease holds for all simply-local
instantiations of a clause.

The second difference concerns the question of whether a decrease is
required for recursive body atoms only, or whether recursion plays
no role.  Simply-acceptability is the only characterisation that
requires a decrease for recursive body atoms only. We attribute this
difference essentially to the explicit use of modes.  Broadly
speaking, modes restrict the data flow of a program in a way that
allows for termination proofs that are inherently {\em modular}.
Therefore one does not explicitly require a decrease for non-recursive
calls, but rather one requires that for the predicate of the
non-recursive call, termination has already been shown
(independently).  To support this explanation, we refer
to~\cite{EBC99}: there left-termination for {\em well moded} programs
is shown, using {\em well-acceptability}.  Well-acceptability requires
a decrease only for recursive body atoms.

The third difference concerns the question of whether the method
relies on (some kind of) models or not. It is not surprising that a
method for showing strong termination cannot rely on models: one
cannot make any assumptions about certain atoms being resolved before
an atom is selected. However, the original methods of showing
termination for input-consuming derivations were also not based on
models~\cite{Sma99,BER99}, and it was remarked that the principle
underlying the use of models in proofs of left-termination cannot be
easily transferred to input termination. By
restricting to simply moded programs and defining a special notion of
model, this was nevertheless achieved.  For a clause $H \la
A_1,\dots,A_n$, assuming that $A_i$ is the selected atom, we exploited
that provided that programs and queries are simply moded, we know that
even though $A_1,\dots,A_{i-1}$ may not be resolved completely,
$A_1,\dots,A_{i-1}\theta$ will be in any ``partial model'' of the
program.

The fourth difference concerns the question of whether proof
obligations are imposed on queries. Delay-recurrence and
simply-acceptability are the characterisations that impose no proof
obligations for queries (except that in the latter case, the query
must be simply moded). The reason is that the restrictions on the
selectability of an atom, which depends on the degree of
instantiation, take the role of such a proof obligation.

The fifth difference concerns the question of whether $\infty$ is
 in the codomain of level mappings. This is the case for
acceptability, fair-boundedness and boundedness. In all three cases,
this allows for excluding {\em unintended atoms\/} and {\em
non-terminating atoms\/} from the termination analysis. For an atom
$A$ with $|A| = \infty$ the proof obligations are trivially satisfied.
Also, the use of $\infty$ allows to achieve full
completeness of the characterisation.

A final difference concerns
the way information on data flow (modes, simply-local models, covers,
Herbrand models) is used in the declarative characterisations. 
For recurrence this is not applicable. Apart from that, in all
except fair-boundedness, such information is used only 
in a ``positive'' way, i.e., ``if \ldots {\em is\/} in the model then
\ldots''. In fair-boundedness, it is also used in a ``negative''
way, namely ``if \ldots {\em is not\/} in the model then \ldots''.
Intuitively, in all characterisation, except fair-boundedness, the
relevant part of the information concerns a characterisation of atoms
that are logical consequences of the program. In fair-boundedness,
it is also relevant the characterisation of atoms that are not
logical consequences, since for those atoms we must ensure finite
failure.

\subsection{From Strong Termination to Bounded Nondeterminism}\label{from-st-to-bn-subsec}
In this subsection, we show inclusions between the introduced classes, i.e., we justify
each arrow in figure~\ref{overview-fig}. 
We first leave aside input termination and local delay
termination, since for these classes, the comparison is much less
clearcut. 

Looking at the four remaining classes from an operational point of
view, we note that strong termination of a program and a query
implies left-termination, which in turn implies
$\exists$-termination, which in turn implies bounded nondeterminism.
Examples~\ref{ex:strong-strictly-left},
\ref{ex:left-strictly-exists} and \ref{ex:exists-strictly-bounded} show
that these implications are strict.

Since the declarative characterisations of those notions are sound and
complete, the same strict inclusions hold among recurrence,
acceptability, fair-boundedness and boundedness.  This allows for
reusing or simplifying termination proofs.

\begin{theorem} \label{c2:r:rec}
Let $P$ be a program and $Q$ a query, $|.|$ an extended level mapping
and $I$ a Herbrand model of $P$. Each of the following statements
strictly implies the statements below it:
\begin{itemize}
\item[{\it (i)\/}] $P$ and $Q$ are recurrent by $|.|$,
\item[{\it (ii)\/}] $P$ and $Q$  are acceptable by $|.|$ and  $I$,
\item[{\it (iii)\/}] $P$ and $Q$ are fair-bounded by $|.|$ and $I$,
\item[{\it (iv)\/}] $P$ and $Q$ are bounded by $|.|$ and $I$.
\end{itemize}
\end{theorem}

In the following example, we show how the above theorem allows for
reuse of termination proofs.

\begin{example}\label{ex:reuse}
In Example~\ref{ex:transp-is-acceptable} we showed that the {\tt TRANSP} program
is acceptable by a level mapping $|.|$ and a model $I$. The proof
obligations of acceptability had to be shown for every clause of the
program.

However, we note that the clauses defining the predicate {\tt member}
are a sub-program which is readily checked to be recurrent by the same
$|.|$. By Theorem~\ref{c2:r:rec}, we conclude that the proof obligations
for clauses defining {\tt member} are satisfied for every Herbrand
model of {\tt TRANSP} and thus in particular for $I$.

We refer the reader to \cite{AP94m} for a collection
of results on reuse of proofs of recurrence to show acceptability, and
on proving acceptability of $P \U P'$ by reusing separate proofs for
$P$ and $P'$.  
\end{example}

Consider now local delay termination. Obviously, it is implied by
strong termination. However, we have observed with the programs and
queries of Examples~\ref{ex:lds-yes-left-no}
and~\ref{ex:ic-or-lds-not-imply-exists} that local delay termination
does not imply left-termination or $\exists$-termination, in general.
These results can be obtained under reasonable assumptions, which, in
particular, rule out deadlock.

The following proposition relates local delay termination with
$\exists$-termination. 

\begin{proposition}\label{local-implies-exists-prop}
Let $P$ and $Q$ be a permutation well moded program and query, and
$|.|$ a moded level mapping.

If $P$ and $Q$ local delay terminate (w.r.t.~$|.|$) then
they $\exists$-terminate.

If $P$ is delay-recurrent by $|.|$ and some Herbrand interpretation
then $P$ and $Q$ are fair-bounded by some extended level mapping and
Herbrand interpretation.
\end{proposition}
\begin{proof}
Since $P$ and $Q$ are permutation well moded, every query $Q'$ in a
  derivation of $P$ and $Q$ is permutation well moded \cite{Sma99t},
  i.e., there exists a permutation $\tilde{Q'}$ of $Q'$ which is well
  moded. By Definition~\ref{well-moded}, the leftmost atom in $\tilde{Q'}$
  is ground in its input positions and hence bounded
  w.r.t.~$|.|$. Consider the selection rule which always selects this
  ``leftmost'' (modulo the permutation) atom. This selection rule is
  local and delay-safe, and it is a standard selection rule (since
  there is always a selected atom). Therefore, local delay termination
  implies $\exists$-termination.

  Concerning the second claim, since fair-boundedness is a complete
  characterisation of $\exists$-termination, we have the conclusion.
\end{proof}

The next proposition relates local delay termination with
left-termination. In this case, programs must be
well moded, not just {\em permutation} well moded. The proof is
similar to the previous one but simpler.

\begin{proposition}\label{local-implies-left-prop}
Let $P$ and $Q$ be a well moded program and query, and $|.|$ a moded
level mapping.

If $P$ and $Q$ local delay terminate (w.r.t.~$|.|$) then
they left-terminate.

If $P$ is delay-recurrent by $|.|$ and some Herbrand interpretation
then $P$ and $Q$ are acceptable by some extended level mapping and
Herbrand interpretation.
\end{proposition}

Marchiori \& Teusink \shortcite{MT99} propose a
program transformation such that the original program is
delay-recurrent iff the transformed program is acceptable. This
transformation allows us to use automated proof methods originally
designed for acceptability for the purpose of showing delay-recurrence.

 Consider now input termination. As before, it is
implied by strong termination. However, as observed in
Examples~\ref{ex:ic-yes-lds-no},~\ref{ex:lds-yes-left-no} and \ref{ex:ic-or-lds-not-imply-exists},
input termination does not imply local delay termination,
left-termination, or $\exists$-termination, in general. Again, these
results can be obtained under reasonable assumptions.

The following proposition relates input termination to
$\exists$-termination. 

\begin{proposition}\label{input-implies-exists-prop}
Let $P$ and $Q$ be a permutation well moded program and query. If $P$
and $Q$ input terminate then they $\exists$-terminate.

Let $P$ and $Q$ be a permutation well and simply moded program and
query. If $P$ is simply-acceptable by some $|.|$ and $I$ then $P$ and
$Q$ are fair-bounded by some extended level mapping and Herbrand
interpretation.
\end{proposition}
\begin{proof}
Since $P$ and $Q$ are permutation well moded, every query $Q'$ in a
  derivation of $P$ and $Q$ is permutation well moded \cite{Sma99t},
  and so $Q'$ contains an atom that is ground in its input
  position. The selection rule $s$ that always selects this atom
  together with all program clauses is an input-consuming selection
  rule, and also a standard selection rule.  Therefore,
  input termination implies universal termination
  w.r.t.~$\{s\}$ and hence $\exists$-termination.

  Concerning the second claim, by Theorem~\ref{input-termination-thm},
  $P$ and $Q$ input terminate. As shown above, this implies
  that they $\exists$-terminate. Since fair-boundedness is a complete
  characterisation of $\exists$-termination, we have the conclusion.
\end{proof}

 The next proposition gives a direct comparison between
input and left-termina\-tion. The proof is similar to the
previous one.

\begin{proposition}\label{input-implies-left-prop}
Let $P$ and $Q$ be a well moded program and query. If $P$ and $Q$
input terminate then they left-terminate.

Let $P$ and $Q$ be a well and simply moded program and query. If $P$
is simply-acceptable by some $|.|$ and $I$ then $P$ and $Q$ are
acceptable by some extended level mapping and Herbrand interpretation.
\end{proposition}

 To relate input termination to local delay termination,
we introduce a notion that relates delay-safe derivations with
input-consuming derivations, based on an a similar concept
from \cite{AL95}.

\begin{definition}\label{implies-matching}
 Let $P$ be a program and $|.|$ a moded level mapping.
 
 We say that $|.|$ {\em implies matching\/}
 (w.r.t.~$|.|$) if for every atom 
$A = p({\mathbf s}, {\mathbf t})$
 bounded w.r.t.~$|.|$ and for every 
$B = p({\mathbf v}, {\mathbf u})$ 
head of a renaming of a clause from $P$ which is variable-disjoint
 with $A$, if $A$ and $B$ unify, then $\mathbf s$ is an instance of
 $\mathbf v$.  
\end{definition}

  Note that, in particular, $|.|$ implies matching if every atom
bounded by $|.|$ is ground 
in its input positions.  

\begin{proposition} \label{ic-imp-ld}\label{input-implies-local-prop}
  Let $P$ and $Q$ be a permutation simply moded program and query, and
  $|.|$ a moded level mapping that implies matching.
 
If $P$ and $Q$ input terminate then they local delay
 terminate (w.r.t.~$|.|$).
\end{proposition}
\begin{proof}
The conclusion follows by
showing that any derivation of $P$ and any permutation simply moded
query $Q'$ via a local delay-safe selection rule (w.r.t.~$|.|$) is
also a derivation via an input-consuming selection rule. So, let $s$
be a local delay-safe selection rule and $Q'$ a permutation
simply-well moded query such that $s$ selects atom $A = p({\mathbf
s},{\mathbf t})$. Then by Definition~\ref{implies-matching}, for each $B =
p({\mathbf v},{\mathbf u})$, head of a renaming of a clause from $P$,
if $A$ and $B$ unify, then $\mathbf s$ is an instance of $\mathbf v$,
i.e. ${\mathbf s} = {\mathbf v}\theta$ for some substitution $\theta$
such that $dom(\theta) \sse \vars(\vect{v})$.  
By \cite[(Apt \& Luitjes, 1995, Corollary 31)]{AL95}, 
this implies that the resolvent of $Q'$ and any clause in
$P$ is again permutation simply moded. Moreover, by applying the
unification algorithm \cite{Apt97}, it is readily checked that, if $A$
and $B$ unify, then $\sigma = \theta \U \{ {\mathbf t}/{\mathbf
u}\theta \}$ is an mgu. Permutation simply-modedness implies that ${\mathbf s}$
and ${\mathbf t}$ are variable-disjoint. Moreover, ${\mathbf s}$ and
${\mathbf v}$ are variable-disjoint. This implies that $\dom(\sigma)
\cap \vars(\vect{s}) = \emptyset$, and so the derivation step is
input-consuming.
 
By repeatedly applying this argument to all queries in the
SLD-derivation of $P$ and $Q$ via $s$, it follows that the derivation
is via some input-consuming selection rule.  
\end{proof}
 
 It remains an open question whether simply-acceptability implies
delay-recurrence under some general hypotheses. The problem with
showing such a result lies in the fact that delay-recurrence is a
sufficient but not necessary condition for local delay termination.
 
\begin{example}\label{ex:open-question}
Consider again the program and the level mapping $|.|$ of
Example~\ref{ex:delay-recursive-incomplete}. 
We have already observed that the program and any
query local delay terminate.

In addition, given the mode $\{ {\tt p(\out)}$, ${\tt q(\inp)}$, ${\tt
r(\inp)} \}$, it is readily checked that the program is
simply moded, and that the level mapping is moded and implies
matching. Also, note that the program is simply-acceptable
by $|.|$ and any simply-local model.

However, this is not sufficient to show that the program is
delay-recurrent, as proved in 
Example~\ref{ex:delay-recursive-incomplete}.  Intuitively, the
problem with showing delay-recurrence lies in the fact that the notion
of cover does not appropriately describe the data flow in this program
given by the modes. 
\end{example}

\subsection{From Bounded Nondeterminism to Strong Termination}
Consider now a program $P$ and a query $Q$ which either do not
universally terminate for a set of selection rules in question, or
simply for which we (or our compiler) fail to {\em prove} termination.
We have
already mentioned that, if $P$ and $Q$ have bounded nondeterminism
then there exists an upper bound for the length of the SLD-refutations
of $P$ and $Q$. If the upper bound is known, then we can syntactically
transform $P$ and $Q$ into an equivalent program and query that
strongly terminate. As shown by Pedreschi \& Ruggieri \shortcite{PR99}, 
such an upper bound is
related to the natural number $k$ of Definition~\ref{c2:def:bq} of
bounded queries. As in our notation for moded atoms, we use boldface
letters to denote vectors of (possibly non-ground) terms.

\begin{definition} \label{c2:def:bntrans}
  Let $P$ be a program and $Q$ a query both bounded by $|.|$ and $I$,
  and let $k \in \nat$. We define $Ter(P)$ as the program such that:
\begin{itemize}
\item for every clause 
$p_0(\mathbf{t}_0) \la
    p_1(\mathbf{t}_1),\ldots,p_n(\mathbf{t}_n)$ 
in $P$, with $n > 0$,
  the clause
\[ 
p_0(\mathbf{t}_0,\mathtt{s(D)}) \la 
  p_1(\mathbf{t}_1,\mathtt{D}),\ldots,p_n(\mathbf{t}_n,\mathtt{D})\]
is in $Ter(P)$, where {\tt D} is a fresh variable,
\item and, for every clause 
$p_0(\mathbf{t}_0)$ in $P$,
the clause
\[  
p_0(\mathbf{t}_0,\_)\la 
\]
is in $Ter(P)$.
\end{itemize}
Also, for the query 
$Q = p_1(\mathbf{t}_1),\ldots,p_n(\mathbf{t}_n)$, 
we define $Ter(Q,k)$ as the query
\[
p_1(\mathbf{t}_1,\mathtt{s}^k(0)),\ldots,p_n(\mathbf{t}_n,\mathtt{s}^k(0)) 
\]
\end{definition}

  The transformed program relates to the original one as shown in
the following theorem.

\begin{theorem}[\cite{PR99,Rug99}]
\label{c2:bn:transfcorr}
Let $P$ be a program and $Q$ a query both bounded by $|.|$ and $I$,
and let $k$ be a given natural number satisfying Definition~\ref{c2:def:bq}. 

Then, for every $n \in \nat$, $Ter(P)$ and $Ter(Q, n)$
strongly terminate.

Moreover, there is a bijection between SLD-refutations of $P$ and $Q$
via a selection rule $s$ and SLD-refutations of $Ter(P)$ and $Ter(Q,
k - 1 )$ via $s$.
\end{theorem}

  The intuitive reading of this result is that the transformed
program and query maintain the same {\em success semantics\/} of the
original program and query. Note that no assumption is
made on the selection rule $s$, i.e.~any selection rule is a
terminating control for the transformed program and query.

\begin{example}\label{ex:oddeven-transformed}
Reconsider the program {\tt ODDEVEN} and the query 
$Q = \mathtt{even(X),\; odd(X)}$ of Example~\ref{ex:exists-strictly-bounded}.
The transformed program $Ter(\mathtt{ODDEVEN})$ is:
\begin{janprogram}
\> even(s(X),s(D)) $\la$ odd(X,D).\\
\> even(0,\_).\\[2ex]
\> odd(s(X),s(D)) $\la$ even(X,D).
\end{janprogram}
and the transformed query $Ter(Q, k - 1)$ for $k = 3$ is {\tt
even(X,s$^2$(0)),odd(X,s$^2$(0))}.  By Theorem~\ref{c2:bn:transfcorr}, the
transformed program and query terminate for {\em any} selection rule,
and the semantics w.r.t.~the original program is preserved modulo the
extra argument added to each predicate.  
\end{example}

The transformations $Ter(P)$ and $Ter(Q, k)$ are of purely theoretical
interest. In practice, one would implement these counters directly
into the compiler/interpreter.  Also, the compiler/interpreter should
include a module that infers an upper bound $k$ automatically.
Approaches to the automatic inference of level mappings and models are
briefly recalled in the next section. Pedreschi \& Ruggieri
\shortcite{PR99} give an example showing how the approach of Decorte
{\em et~al.}\ \shortcite{DDV99} could be rephrased to infer
boundedness.

\section{Related Work}\label{related-sec}
The survey on termination of logic programs by De Schreye \& Decorte
\shortcite{SD94} covers most work in this area until 1994. The authors
distinguish three types of approaches: the ones that express necessary
and sufficient conditions for termination, the ones that provide
decidable {\em sufficient\/} conditions, and the ones that prove
decidability or undecidability for subclasses of programs and queries.
Under this classification, our survey falls in the first type.  In the
following, we mainly mention works published since 1994. We group the
works according to the main focus or angle they take.

\subsection{Other Characterisations of Left-termination}

Apt \& Pedreschi \shortcite{AP94m} refined acceptability to make the
method {\em modular}.  Here, modularity means that the termination
proof for a program $P \U P'$ can be obtained from separate
termination proofs for $P$ and $P'$. Also, in \cite{AMP94},
acceptability is extended to reason on first-order built-in's of
Prolog.

Etalle {\em et~al.}~\index{Etalle} \shortcite{EBC99} propose a
refinement of acceptability (called {\em well-acceptability}) for
well moded programs and queries.  The requirement of well-modedness
simplifies proofs of acceptability. On the one hand, no proof
obligation is imposed on the \emph{queries}.  On the other hand, the
decrease of the level mapping is now required only from the head to
the mutually recursive clause body atoms. It is interesting to observe
that the definition of well-acceptability is then very close to
simply-acceptability. Actually, well-modedness of a program and a query
implies that atoms selected by the LD selection rule are ground in
their input positions, hence a derivation via the LD selection rule is
input-consuming.

Serebrenik and De Schreye \shortcite{SS01} show that, when restricting
to well moded programs and queries and moded level mappings (they call
them {\em output-independent}), acceptability can be generalised by
having any well-founded ordering, not necessarily $\nat$, as co-domain
of level mappings. This simplifies the proof of programs where complex
level mappings may be required.

Also, a characterisation of acceptability in the context of metric
spaces was provided by Hitzler \& Seda \shortcite{HS99}.

Alternative characterisations of left-termination consider proof
obligations on generalised level mappings and thus on possibly
non-ground instances of clauses and queries.  Bossi {\em
et~al.}~\shortcite{BCF94} provide sufficient and necessary conditions
that involve: (1) generalised level mappings (with an arbitrary 
well-founded ordering as the codomain) that do not increase w.r.t.\
substitutions; (2) a specification (\pre, \post), with \pre, \post
\sse $Atom_L$, which is intended to characterise call patterns (\pre)
and correct instances (\post) of atomic queries.  Call patterns
provide information on the structure of selected atoms, while correct
instances provide information on data flow. The method has the
advantage of reasoning both on
termination and on partial correctness within the same framework.
However, proof obligations are not well suited for {\em paper \& pencil}
proofs, since they require to reason on the strongly connected
components of a graph abstracting the flow of control of the program
under consideration. An adaption of acceptability to total correctness
is presented in \cite{PR95v}.  Also, we mention the works of Bronsard
{\em et~al.}~\shortcite{BLR92} and Deransart \&
Ma{\l}uszy{\'n}ski~\shortcite{DM93}, which rely on partial correctness
or typing information to characterise call patterns.  Deransart \&
Ma{\l}uszy{\'n}ski generalise the proof obligations on the
left-to-right order of the LD selection rule to any acyclic ordering of
body atoms. Another characterisation of left-termination particularly
suited for automation is due to De Schreye {\em
et~al.}~\shortcite{SVB92,DDV99}. Their notion is similar to the one of
Bossi {\em et~al.}, but it uses: (1) generalised level mappings that are
constant w.r.t.\ substitution (called {\em rigid} level mappings); (2)
a pair (\pre, \post), with \pre, \post \sse $Atom_L$, where \post\ is
a model of the program and \pre\ is a characterisation of call
patterns computed using abstract interpretation.

A generalisation of the definition of left-termination considers a
program together with a {\em set} of queries \cite{SVB92,BCF94}, while
we considered a program and a single query.  We say that a
program $P$ and a set of queries $\cal Q$
left-terminate if every derivation for $P$ and any $Q \in \cal Q$ via
the leftmost selection rule is finite. The benefit of such a
definition consists of having just one single proof of termination for
a set of queries rather than a set of proofs, one for each query in
the set.  However, we observe that in our examples on acceptability,
proofs can easily be generalised to a set of queries.  For instance,
for a level mapping such that $|${\tt p($t$)}$| = |t|$, it is
immediate to conclude that all queries {\tt p($T$)}, where $T$ is a
list, are acceptable.  Conversely, is it the case that if $P$ and
${\cal Q}$ left-terminate then $P$ and any $Q \in \cal Q$ are
acceptable by a same $|.|$ and $I$? The answer is affirmative. In
fact, from the proof of the Completeness Theorem~\ref{c2:acc:tcor}
\cite[(Ruggieri, 1999, Theorem~2.3.20)]{Rug99}, 
if $P$ and $Q$ left-terminate
then they are acceptable by a level mapping $|.|_P$ and a Herbrand
model $I_P$ that {\em only} depend on $P$. This implies that every $Q
\in \cal Q$ is acceptable by $|.|_P$ and $I_P$.  In conclusion,
acceptability by $|.|_P$ and $I_P$ precisely characterises the maximal
set $\cal Q$ such that $P$ and $\cal Q$ left-terminate.

Finally, instead of considering left-termination of a program $P$ and
a query $Q$, one may be
interested in proving left-termination of some permutation $P'$ and
$Q'$ of them. A permutation of $P$ ($Q$) is any program (query)
obtained by reordering clause body atoms in $P$ ($Q$).
This notion is called $\sigma$-termination in \cite{HM01}, where a
system for automatic inference is presented.  $\sigma$-termination is
strictly weaker than left-termination, and strictly stronger than
$\exists$-termination (e.g., program {\tt PRODCONS} in
figure~\ref{prodcons-prog} and {\tt system($n$)}, with $n \in \nat$,
$\exists$-terminate but do not $\sigma$-terminate).

\subsection{Writing Left-Terminating Programs}
There are also works that are not directly concerned with proving an
existing program left-terminating, but rather with heuristics and
transformations that help write left-terminating programs. 

Hoarau \& Mesnard \shortcite{HM98} studied inferring and
compiling termination for (constraint) logic programs. {\em Inferring}
termination means inferring a set of queries for which a program
``potentially'' terminates, that is to say, it terminates after
possible reordering of atoms. This phase uses abstract interpretation and the
Boolean $\mu$-calculus. {\em Compiling} termination means reordering
the body atoms so that the program terminates. The method is
implemented.

Neumerkel \& Mesnard \shortcite{NM99} studied the problem of localising
and explaining reasons for nontermination in a logic programs. The
work aims at assisting programmers in writing terminating programs and
helping them to {\em understand} why their program does not
terminate. The method has been implemented and is intended as a
debugging tool, in particular for beginners (it has been used for
teaching purposes).  The idea is to localise a fragment of a program
that is in itself already non-terminating, and hence constitutes an
explanation for non-termination of the whole program.

\subsection{Transformational Approaches}
It is possible to investigate termination of logic programs by
transforming them to some other formal system. If the transformation
preserves termination, one can resort to the compendium of techniques
of those formal systems for the purpose of proving termination of
the original logic program.

Baudinet \shortcite{Bau92} considered transforming logic programs
into functional programs.  Termination of the transformed programs can
then be studied by structural induction. Her approach covers general
logic programs, existential termination and the effects of Prolog cut.

There is a considerable amount of literature on transforming logic
programs to term rewriting systems (TRSs), which are perhaps the
generic formalism for studying termination as such.  It is very common
in these transformational approaches to use modes. The intuitive idea
is usually that the input of an atom has to rewrite into the output of
that atom. Most of those works assume the left-to-right selection
rule. One valuable exception is due to Krishna Rao {\em
  et~al.}~\shortcite{KRKS98}, where termination is considered
w.r.t.~selection rules that respect a producer-consumer relation among
variables in clauses. Such a producer-consumer relation is formalised
with an extension of the notion of well-modedness. The approach improves
over the original proposal of the authors \cite{KKS92}, where
the LD selection rule was assumed.

The approach by Aguzzi \& Modigliani \shortcite{AguMod93} takes into
account that logic programs can be used in several modes, even within
the same run of a program. Moreover, the approach is able to handle
{\em local} variables, i.e., variables occurring only in a clause body
but not in the head. Such variables model what is
sometimes called {\em sideways information passing}. One remarkable
property of the transformation is that it provides a characterisation
of termination, albeit only for the limited class of {\em input
  driven\/} logic programs \cite{AE93}.  So for this limited class, a
program terminates if and only if the corresponding TRS terminates.

Ganzinger \& Waldmann \shortcite{GW92} proposed a transformation of
logic programs into {\em conditional} TRSs. In such TRSs, the rules
have the form $u_1\to v_1,\dots,u_n\to v_n \Rightarrow
s\to t$, which is to be read as ``if each $u_i$ rewrites to
$v_i$, then $s$ rewrites to $t$''.  Well moded logic program clauses
are transformed into such rules, where there is a correspondence
between each $u_i$ and the input of the $i$th body atom, each $v_i$
and the output of the $i$th body atom, $s$ and the input of the head,
and $t$ and the output of the head.  The method improves over
\cite{KKS92} in applicability and simplicity.

Marchiori \shortcite{MarM94} improves over the transformations of
\cite{AguMod93} and \cite{GW92} by adopting enhanced methods to detect
unification-freeness, i.e. situations where unification (used by
SLD-resolution) boils down to matching (used by TRS operational
semantics).  Another contribution lies in the fact that the
transformation proposed is modular, i.e., it considers each clause in
isolation.  

More recently, Arts \shortcite{Art97} investigated a new termination
method for TRSs called \emph{innermost normalisation} and applied it
also to TRSs obtained by transforming well moded logic programs. The
technique improves over \cite{KKS92}.

\subsection{Dynamic Selection Rules}
By {\em dynamic\/} selection rules we mean those rules where selection
depends on the degree of instantiation of atoms at run-time.  Second
generation logic languages adopt dynamic selection rules as control
primitives. We mention here delay
declarations, input-consuming derivations and guarded clauses.

Apt \& Luitjes \shortcite{AL95} consider deterministic programs, i.e.,
programs where for each selectable atom (according to the delay
declarations), there is at most one clause head unifiable with it. For
such programs, the existence of one successful derivation implies that
all derivations are finite. Apt \& Luitjes also give conditions for
the termination of {\tt append}, but these are ad-hoc and do not
address the general problem.

L{\"u}ttringhaus-Kappel~\shortcite{Lut93} proposes a method for
generating control (delay declarations) automatically.  The method
finds {\em acceptable} delay declarations, ensuring that the most
general selectable atoms have finite SLD-trees. What is required
however are {\em safe} delay declarations, ensuring that {\em
instances} of most general selectable atoms have finite SLD-trees. A
{\em safe} program is a program for which every acceptable delay
declaration is safe.  L{\"u}ttringhaus-Kappel states that all programs
he has considered are safe, but gives no hint as to how this might be
shown in general.  This work is hence not about {\em proving}
termination.  In some cases, the delay declarations that are generated
require an argument of an atom to be a list before that atom can be
selected. This is similar to requiring the atom to be bounded, i.e. to
the approach of \cite{MT99,MK97} and of section~\ref{local-sec}.

Naish \shortcite{Nai92c} considers delay declarations that test for
partial instantiation of certain predicate arguments. Such delay
declarations implicitly ensure input-consu\-ming derivations. He gives
good intuitive explanations about possible causes of loops,
essentially {\em circular modes} and {\em speculative output
bindings}. The first cause (see Example~\ref{circular-ex}) can be
eliminated by requiring programs to be {\em permutation
nicely\footnote{A slightly more general notion than permutation {\em
simply}-modedness.} moded}.  Speculative output bindings are indeed a
good explanation for the fact that $\tt permute(\out,\inp)$ (see
Example~\ref{ex:ic-strictly-lds}) does not input terminate. Naish then
makes the additional assumption that the selection rule always selects
the leftmost selectable atom, and proposes to put recursive calls last
in clause bodies. Effectively, this guarantees that the recursive
calls are {\em ground} in their input positions, which goes beyond
assuming input-consuming derivations.

Naish's proposal has been formalised and refined by Smaus {\em
et~al.}~\shortcite{SHK01}. They consider atoms that may loop when
called with insufficient input, or in other words, atoms for which
assuming input-consuming derivations is insufficient to guarantee
termination. It is proposed to place such atoms sufficiently late; all
producers of input for such atoms must occur textually earlier.
Effectively, this is an assumption about the selection rule that lies
between input-consuming derivations and local delay-safe derivations.

Our characterisation of input termination only requires (permutation)
simply moded programs and queries. The first sound but incomplete
characterisation of \cite{Sma99} assumed well and nicely moded
programs. It was then found that the condition of well-modedness could
easily be lifted~\cite{BER99}.  It was only by restricting to {\em
  simply} moded programs that one could give a characterisation that
is also complete. This means of course that the method of \cite{BER99}
does not subsume the method of \cite{BERS01} surveyed here, but
nevertheless, we believe that the fact that the characterisation is
complete is more important.  Input-consuming derivations can be
ensured in existing systems using {\em delay declarations} such as
provided by G{\"o}del \cite{HL94} or SICStus \cite{sicstus}. This is
shown in \cite{BER00,BERS01,Sma99t}.

The definition of input-consuming derivations has a certain
resemblance with derivations in the parallel logic language of {\em
  (Flat) Guarded Horn Clauses} \cite{Ued88}. In (F)GHC, an atom and
clause may be resolved only if the atom is an instance of the clause
head, and a test ({\em guard}) on clause selectability is satisfied.
Termination of GHC programs was studied by Krishna Rao {\em
  et~al.}~\shortcite{KKS97} by transforming them into TRSs. 

Pedreschi \& Ruggieri \shortcite{PR99fail} characterised a
class of programs with guards and queries that have no
failed derivation.  For those programs, termination for one selection
rule implies termination (with success) for all selection rules. This
situation has been previously described as saying that a program does
not make speculative bindings~\cite{SHK01}.  The approach by Pedreschi
\& Ruggieri is an improvement w.r.t.~the latter one, since what might
be called ``shallow'' failure does not count as failure. For example,
the program \texttt{QUICKSORT} is considered failure-free in the
approach of Pedreschi \& Ruggieri.

\subsection{$\exists$-Termination and Bounded Nondeterminism}

Concerning termination w.r.t. fair selection rules, i.e.,
$\exists$-termination, we are aware only of the works of Gori
\shortcite{Gor00} and McPhee \shortcite{McPhee00}. Gori proposes an
automatic system based on abstract interpretation analysis that infers
$\exists$-termination.  McPhee proposes the notion of {\em prioritised
  fair selection rules}, where atoms that are known to terminate are
selected first, with the aim of improving efficiency of fair selection
rules. He adopts the automatic test of Lindenstrauss \& Sagiv 
\shortcite{LS97} to infer
(left-)termination, but, in principle, the idea applies to any
automatic termination inference system.

Concerning bounded nondeterminism, Martin \& King \shortcite{MK97}
define a transformation for G{\"o}del programs, which shares with the
transformation of Definition~\ref{c2:def:bntrans} 
the idea of not following derivations longer than a certain length.
However, they rely on sufficient conditions for inferring the length
of refutations, namely termination via a class of selection rules
called {\em semilocal}. Their transformation adds run-time overhead,
since the maximum length is computed at run-time. On the other hand, a
run-time analysis is potentially able to generate more precise upper
bounds than our static transformation, and thus to cut more
unsuccessful branches. Also, the idea of pruning SLD-derivations at
run-time is common to the research area of loop checking \cite{BAK91}.

Sufficient (semi-)automatic methods to approximate the number of
computed instances by means of lower and upper bounds have been
studied in the context of cost analysis of logic programs \cite{DL93}
and of cardinality analysis of Prolog programs \cite{BCMV94}. Of
course, if $\infty$ is a lower bound to the number of computed
instances of $P$ and $Q$ then they do not have bounded nondeterminism.
Dually, if $n \in N$ is an upper bound then $P$ and $Q$ have bounded
nondeterminism. In this case, however, we are still left with the
problem of determining a depth of the SLD-tree that includes all the
refutations.

\subsection{Automatic Termination Inference}\label{subsec:automatic}
On a theoretical level, the problem of deciding whether a program
belongs to one of the classes studied in this article is undecidable.
This was formally shown by Bezem \shortcite{Bez93} for recurrence, and
by Ruggieri \shortcite{Rug99} for acceptability, fair-boundedness and
boundedness. On a practical level, however, many methods have been
proposed to infer (usually: left-) termination automatically.  This
research stream is currently very active, and some efficient tools
are already integrated in compilers.

A challenging topic of the research in automatic termination inference
consists in finding standard forms of level mappings and models,
so that the solution of the resulting proof obligations can be
reduced to known problems for which efficient algorithms exist
\cite{BCF94,BK97,DDF93,Plu90,vG91}.

As an example, we mention the detailed account of automatic
termination analysis by Decorte {\em et~al.}\ \shortcite{DDV99}.
The main idea is as follows. Termination analysis is
parametrised by several factors, such as the choice of modes and 
level mappings. In practice, these are usually inferred using abstract
interpretation techniques. This is often not very efficient.
Therefore Decorte {\em et~al.} propose to encode
all those parameters and the conditions that have to hold for
them as constraints. So for example, there are constraint
variables for each weighting parameter used in the definition of
(semi-) linear norms and level mappings. To show termination of the
analysed program, one has to find a solution to the constraint system.

Lindenstrauss \& Sagiv \shortcite{LS97} developed the system {\em
TermiLog\/} for checking termination. They use linear
norms, (monotonicity and equality) constraint inference and the
termination test of Sagiv \shortcite{Sag91}, originally designed for
Datalog programs. The implementation of the static termination
analysis algorithm of the Mercury system ~\cite{SSS97} exploits mode
and type information provided by the programmer. Speirs {\em et~al.}
claim a better performance than the {\em TermiLog\/} system in the
average case.  The implementation of fair selection rules has been
announced for future releases of Mercury. Codish \index{Codish} \&
Taboch \index{Taboch} \shortcite{CT99} proposed a formal semantics
basis that facilitates abstract interpretation for inferring
left-termination.

Recently, Mesnard {\em et~al.}~\shortcite{Mesnard00} developed the cTI
system for {\em bottom-up} left-termination inference of logic
programs. Bottom-up refers to the use of abstract interpretation based
fixpoint computations whose output is a set of queries for which the
system infers termination. The results show that, on several benchmark
programs, the sets of queries inferred by cTI strictly include the set
of queries for which the top-down methods of \cite{DDV99,LS97,SSS97}
can show termination.

Finally, we recall the approach by St{\"a}rk~\shortcite{Sta98} to prove
both termination and partial correctness together. His system, called
LPTP, is implemented in Prolog and consists of an interactive theorem
prover able to prove termination and correctness of Prolog programs
with negation, arithmetic built-in's and meta-predicates such as {\tt
  call}. The formal theory underlying LPTP is an
inductive extension of pure Prolog programs that allows to express
modes and types of predicates.

\subsection{Extensions of Pure Logic Programming}
In this article, we have assumed the standard definition of
SLD-derivations for definite logic programs. We now briefly discuss
termination of alternative or generalised execution models of logic
programs.

A declarative characterisation of strong termination for {\em general}
logic programs and queries (i.e., with negation) was proposed by Apt
\& Bezem \shortcite{AB91}.  The execution model assumed is {\em SLDNF}
resolution with a {\em safe} (not to be confused with {\em delay-safe}
\cite{MT99}) selection rule, meaning that negative literals can be
selected only if they are ground. Also, we mention the bottom-up approach of
Balbiani~\shortcite{Bal91}, where an operator $T_P$ is provided such
that its ordinal closure coincides with those ground atoms $A$ such
that $P$ and $A$ strongly terminate.

Apt \& Pedreschi \shortcite{AP93r} have generalised acceptability to
reason on programs with negation under SLDNF resolution. The
characterisation is sound. Also, it is complete for safe selection
rules. Marchiori \shortcite{Mar95c} proposes a modification of
acceptability to reason on programs with Chan's
constructive negation resolution.

Termination of {\em abductive} logic programs has been studied by Verbaeten
\shortcite{Ver99}. The execution model of abductive logic programs,
called {\em SLDNFA} resolution, extends SLDNF resolution. 
Just as for Apt \& Bezem \shortcite{AB91}, the selection rule is an 
arbitrary safe one, but 
the definition of safe is weaker in this context. Essentially,
SLDNFA resolution behaves worse than SLDNF resolution
w.r.t.~termination, which is why the conditions given by Apt \& Bezem 
\shortcite{AB91}
have to be strengthened. Finally, we point out that the conditions
given are sufficient but not necessary.

{\em Tabled} logic programming is particularly interesting in the
context of termination analysis since tabling improves the termination
behaviour of a logic program, compared to ordinary execution. The
works we discuss in the following take advantage of this, i.e., they can show
termination in interesting cases where ordinary execution does not
terminate. They assume tabled execution based on the left-to-right
selection rule.

A declarative characterisation of tabled termination has been given by
Decorte {\em et~al.}~\shortcite{DDLMS98}. To automate termination
proofs of tabled logic programs, this work has been combined by
Verbaeten \& De Schreye \shortcite{VD01} with the constraint-based
approach to proving left-termination automatically, discussed above
\cite{DDV99}. Verbaeten {\em et~al.}~\shortcite{VSD01} have studied
termination of programs using a mix of tabled and ordinary execution.

Concerning {\em constraint} logic programming (CLP), Colussi 
{\em et al.}~\shortcite{CMM95} first proposed a necessary and sufficient condition for
left-termination, inspired by the method of Floyd for termination of
flowchart programs. Their method consists of assigning a data flow
graph to a program, and then to state conditions to prevent the
program to enter an infinite loop in the graph.

Also, Ruggieri~\shortcite{Rug97t} proposed an extension of
acceptability that is sound and complete for {\em ideal} CLP
languages. A CLP language is ideal if its constraint solver, the
procedure used to test consistency of constraints, returns $true$ on a
consistent constraint and $false$ on an inconsistent one. In
contrast, a non-ideal constraint solver may return $unknown$ if it is
unable to determine (in)consistency.  An example of non-ideal
CLP language is the CLP(${\cal R}$) system, for which Ruggieri
proposes additional proof obligations (based on a notion of modes) to
acceptability in order to obtain a sound characterisation of
left-termination.

Mesnard~\shortcite{Mes96} provides sufficient termination conditions
based on approximation techniques and boolean $\mu$-calculus, with the
aim of {\em inferring\/} a class of left-terminating CLP queries.  The
approach has been refined and implemented by Hoarau and
Mesnard~\shortcite{HM98}.

\section{Conclusion}\label{conc-sec}
In this article, we have surveyed six different classes of terminating
logic programs and queries.  For each class, we have provided a sound
declarative characterisation of termination. Except for local delay
termination, this characterisation was also complete.  We have offered
a unified view of those classes allowing for non-trivial formal
comparisons.

In subsection~\ref{characterisations-subsec}, we have compared the
different characterisations w.r.t.~certain technical details with the
aim of  understanding the role each technical detail plays.

In subsection~\ref{from-st-to-bn-subsec}, we have compared
the classes themselves. The inclusion relations among them are 
summarised in the hierarchy of figure~\ref{overview-fig}.
Intuitively, as the assumptions about the selection rule become
stronger, the proof obligations about programs become weaker. 

One may ask: in how far is such a hierarchy ad-hoc, and could other
classes be considered? We believe that the interest in strong
termination, $\exists$-termination and bounded nondeterminism is
evident because they are cornerstones of the whole spectrum of
classes. The interest in left-termination is motivated by the fact
that the standard selection rule of Prolog is assumed.

The interest in input termination and local delay termination is more
arguable. We cannot claim that there are no other interesting classes
in the surroundings of those two classes. Nevertheless, we believe
that the distinction between input-consuming and local delay-safe
selection rules captures an important difference among dynamic selection
rules: requiring derivations to be input-consuming can be considered a 
reasonable minimum requirement to ensure termination, as we have
argued that only very simple or contrived programs strongly
terminate. In particular, the selection rule does not allow for
methods showing termination that rely on boundedness of the selected
atom. At the time of the selection, the depth of the SLD tree of an
atom is not determined (by the atom itself). In contrast, local
delay-safe selection rules require that the selected atom is bounded,
and thus the depth of the SLD tree of an atom is determined.

We thus hope that we have captured much of the essence of the effect
different choices of selection rules have on termination. This should
be a step towards a possible automatic choice of selection rule and
thus towards realising Kowalski's ideal.

\bibliography{class}

\end{document}